\documentclass[aps,prd,twocolumn,showpacs,superscriptaddress,nofootinbib]{revtex4}

\usepackage{amsmath}   
\usepackage{graphicx}  
\usepackage{xspace}
\usepackage{units}
\usepackage{lscape}

\usepackage{hyperref}

\newcommand{\minerva}{MINERvA\xspace}

\newcommand{\sizecheck}{0} 

\newif\ifpdf
\ifx\pdfoutput\undefined
   \pdffalse
\else
   \pdfoutput=1
   \pdftrue
\fi
\ifpdf
   \usepackage{graphicx}
   \usepackage{epstopdf}
\else
   \usepackage{graphicx}
\fi

\begin{document}

\title{Neutron measurements from antineutrino hydrocarbon reactions}




\newcommand{\deceased}{Deceased}


\newcommand{\Rutgers}{Rutgers, The State University of New Jersey, Piscataway, New Jersey 08854, USA}
\newcommand{\Hampton}{Hampton University, Dept. of Physics, Hampton, VA 23668, USA}
\newcommand{\Dortmund}{Institute of Physics, Dortmund University, 44221, Germany }
\newcommand{\Otterbein}{Department of Physics, Otterbein University, 1 South Grove Street, Westerville, OH, 43081 USA}
\newcommand{\JMU}{James Madison University, Harrisonburg, Virginia 22807, USA}
\newcommand{\Florida}{University of Florida, Department of Physics, Gainesville, FL 32611}
\newcommand{\UCIrvine}{Department of Physics and Astronomy, University of California, Irvine, Irvine, California 92697-4575, USA}
\newcommand{\CBPF}{Centro Brasileiro de Pesquisas F\'{i}sicas, Rua Dr. Xavier Sigaud 150, Urca, Rio de Janeiro, Rio de Janeiro, 22290-180, Brazil}
\newcommand{\PUCP}{Secci\'{o}n F\'{i}sica, Departamento de Ciencias, Pontificia Universidad Cat\'{o}lica del Per\'{u}, Apartado 1761, Lima, Per\'{u}}
\newcommand{\INRM}{Institute for Nuclear Research of the Russian Academy of Sciences, 117312 Moscow, Russia}
\newcommand{\Jlab}{Jefferson Lab, 12000 Jefferson Avenue, Newport News, VA 23606, USA}
\newcommand{\Pittsburgh}{Department of Physics and Astronomy, University of Pittsburgh, Pittsburgh, Pennsylvania 15260, USA}
\newcommand{\Guanajuato}{Campus Le\'{o}n y Campus Guanajuato, Universidad de Guanajuato, Lascurain de Retana No. 5, Colonia Centro, Guanajuato 36000, Guanajuato M\'{e}xico.}
\newcommand{\Athens}{Department of Physics, University of Athens, GR-15771 Athens, Greece}
\newcommand{\Tufts}{Physics Department, Tufts University, Medford, Massachusetts 02155, USA}
\newcommand{\WM}{Department of Physics, College of William \& Mary, Williamsburg, Virginia 23187, USA}
\newcommand{\FNAL}{Fermi National Accelerator Laboratory, Batavia, Illinois 60510, USA}
\newcommand{\Purdue}{Department of Chemistry and Physics, Purdue University Calumet, Hammond, Indiana 46323, USA}
\newcommand{\MCLA}{Massachusetts College of Liberal Arts, 375 Church Street, North Adams, MA 01247}
\newcommand{\UMD}{Department of Physics, University of Minnesota -- Duluth, Duluth, Minnesota 55812, USA}
\newcommand{\Northwestern}{Northwestern University, Evanston, Illinois 60208}
\newcommand{\Mississippi}{University of Mississippi, University, Mississippi 38677, USA}
\newcommand{\UNI}{Universidad Nacional de Ingenier\'{i}a, Apartado 31139, Lima, Per\'{u}}
\newcommand{\Rochester}{University of Rochester, Rochester, New York 14627 USA}
\newcommand{\Austin}{Department of Physics, University of Texas, 1 University Station, Austin, Texas 78712, USA}
\newcommand{\USM}{Departamento de F\'{i}sica, Universidad T\'{e}cnica Federico Santa Mar\'{i}a, Avenida Espa\~{n}a 1680 Casilla 110-V, Valpara\'{i}so, Chile}
\newcommand{\Geneva}{University of Geneva, 1211 Geneva 4, Switzerland}
\newcommand{\Chicago}{Enrico Fermi Institute, University of Chicago, Chicago, IL 60637 USA}
\newcommand{\OregonState}{Department of Physics, Oregon State University, Corvallis, Oregon 97331, USA}
\newcommand{\oxford}{Oxford University, Department of Physics, Oxford, United Kingdom}
\newcommand{\umiss}{University of Mississippi, Oxford, Mississippi 38677, USA}
\newcommand{\upenn}{University of Pennsylvania, Philadelphia, Pennsylvania 19104, USA}
\newcommand{\AMU}{AMU Campus, Aligarh, Uttar Pradesh 202001, India}
\newcommand{\wroclaw}{University of Wroclaw, plac Uniwersytecki 1, 50-137 Wrocław, Poland}
\newcommand{\Mohali}{Indian Institute of Science Education and Research - Mohali, Knowledge city, Sector 81, SAS Nagar, Manauli PO 140306}
\newcommand{\chrismarshallThanks}{now at Lawrence Berkeley National Lab, Berkeley, CA 94720}
\newcommand{\cpatrickThanks}{now at University College London, London WC1E 6BT, UK}
\newcommand{\jwolcottThanks}{now at Tufts University, Medford, MA 02155, USA}
\newcommand{\melkinsThanks}{now at Iowa State University, Ames, IA 50011, USA }
\newcommand{\joelmousseauThanks}{now at University of Michigan, Ann Arbor, MI 48109, USA}
\newcommand{\leisticoThanks}{now at University of Illinois, Champaign,  IL 61820, USA}
\newcommand{\lovleinThanks}{now at University of Utah, Salt Lake City,  UT 84112, USA}
\newcommand{\albinThanks}{NSF supported Soudan Outreach Undergraduate
  Intern}

\author{M.~Elkins}\thanks{\melkinsThanks}                     \affiliation{\UMD}
\author{T.~Cai}                           \affiliation{\Rochester}
\author{J.~Chaves}                           \affiliation{\upenn}
\author{J.~Kleykamp}                      \affiliation{\Rochester}

\author{F.~Akbar}                         \affiliation{\AMU}
\author{L.~Albin}                           \affiliation{\UMD}
\author{L.~Aliaga}                        \affiliation{\WM}  \affiliation{\PUCP}
\author{D.A.~Andrade}                     \affiliation{\Guanajuato}
\author{M.V.~Ascencio}                   \affiliation{\PUCP}
\author{A.~Bashyal}                       \affiliation{\OregonState}
\author{L.~Bellantoni}                    \affiliation{\FNAL}
\author{A.~Bercellie}                     \affiliation{\Rochester}
\author{M.~Betancourt}                    \affiliation{\FNAL}
\author{A.~Bodek}                         \affiliation{\Rochester}
\author{A.~Bravar}                        \affiliation{\Geneva}
\author{H.~Budd}                          \affiliation{\Rochester}
\author{G.~Caceres}                       \affiliation{\CBPF}
\author{M.F.~Carneiro}                    \affiliation{\OregonState}
\author{D.~Coplowe}                       \affiliation{\oxford}
\author{H.~da~Motta}                      \affiliation{\CBPF}
\author{S.A.~Dytman}                      \affiliation{\Pittsburgh}
\author{G.A.~D\'{i}az~}                   \affiliation{\Rochester}  \affiliation{\PUCP}
\author{J.~Felix}                         \affiliation{\Guanajuato}
\author{L.~Fields}                        \affiliation{\FNAL} \affiliation{\Northwestern}
\author{A.~Filkins}                       \affiliation{\WM}
\author{R.~Fine}                          \affiliation{\Rochester}
\author{N.~Fiza}                          \affiliation{\Mohali}
\author{A.M.~Gago}                        \affiliation{\PUCP}
\author{R.~Galindo}                        \affiliation{\USM}
\author{A.~Ghosh}                         \affiliation{\USM}  \affiliation{\CBPF}
\author{R.~Gran}                          \affiliation{\UMD}
\author{J.Y.~Han}                         \affiliation{\Pittsburgh}
\author{A.~Habig}           \affiliation{\UMD}
\author{D.A.~Harris}                      \affiliation{\FNAL}
\author{S.~Henry}                         \affiliation{\Rochester}
\author{S.~Jena}                          \affiliation{\Mohali}
\author{D.~Jena}                           \affiliation{\FNAL}
\author{M.~Kordosky}                      \affiliation{\WM}
\author{D.~Last}                          \affiliation{\upenn}
\author{T.~Le}                            \affiliation{\Tufts} \affiliation{\Rutgers}
\author{J.R.~Leistico}  \affiliation{\UMD}
\author{A.G.~Lopez}  \affiliation{\UMD} 
\author{A.~Lovlein}  \affiliation{\UMD} 
\author{X.-G.~Lu}                         \affiliation{\oxford}
\author{E.~Maher}                         \affiliation{\MCLA}
\author{S.~Manly}                         \affiliation{\Rochester}
\author{W.A.~Mann}                        \affiliation{\Tufts}
\author{C.M.~Marshall}\thanks{\chrismarshallThanks}  \affiliation{\Rochester}
\author{C.~Mauger}                        \affiliation{\upenn}
\author{A.M.~McGowan}                     \affiliation{\Rochester}
\author{K.S.~McFarland}                   \affiliation{\Rochester}  \affiliation{\FNAL}
\author{B.~Messerly}                      \affiliation{\Pittsburgh}
\author{J.~Miller}                        \affiliation{\USM}
\author{J.G.~Morf\'{i}n}                  \affiliation{\FNAL}
\author{J.~Mousseau}\thanks{\joelmousseauThanks}  \affiliation{\Florida}
\author{D.~Naples}                        \affiliation{\Pittsburgh}
\author{J.K.~Nelson}                      \affiliation{\WM}
\author{C.~Nguyen}                       \affiliation{\Florida}
\author{A.~Norrick}                       \affiliation{\FNAL}  \affiliation{\WM}
\author{Nuruzzaman}                       \affiliation{\Rutgers}  \affiliation{\USM}
\author{A.~Olivier}                       \affiliation{\Rochester}
\author{V.~Paolone}                       \affiliation{\Pittsburgh}
\author{G.N.~Perdue}                      \affiliation{\FNAL}  \affiliation{\Rochester}
\author{M.A.~Ram\'{i}rez}                 \affiliation{\Guanajuato}
\author{R.D.~Ransome}                     \affiliation{\Rutgers}
\author{H.~Ray}                           \affiliation{\Florida}
\author{D.~Rimal}                         \affiliation{\Florida}
\author{P.A.~Rodrigues}          \affiliation{\oxford}  \affiliation{\umiss}     \affiliation{\Rochester}
\author{D.~Ruterbories}                   \affiliation{\Rochester}
\author{H.~Schellman}                     \affiliation{\OregonState}  \affiliation{\Northwestern}
\author{C.J.~Solano~Salinas}              \affiliation{\UNI}
\author{H.~Su}                            \affiliation{\Pittsburgh}
\author{V.S.~Syrotenko}                   \affiliation{\Tufts}
\author{S.~S\'{a}nchez~Falero}            \affiliation{\PUCP}
\author{E.~Valencia}                      \affiliation{\WM}  \affiliation{\Guanajuato}
\author{J.~Wolcott}\thanks{\jwolcottThanks}  \affiliation{\Rochester}
\author{B.~Yaeggy}                        \affiliation{\USM}


%
\collaboration{\minerva  Collaboration}\ \noaffiliation

\date{\today}

\begin{abstract}
Charged-current antineutrino interactions on hydrocarbon scintillator
in the MINERvA detector are used to study activity from their
final-state neutrons. 
To ensure that  most of the neutrons are from
the primary interaction, rather than hadronic reinteractions in the
detector, the sample is limited to momentum transfers below 0.8 GeV/c.
From 16,129 interactions, 15,246 neutral particle candidates are
observed.
The reference simulation predicts 64\% of these candidates
are due to neutrons from the antineutrino interaction directly, but also
overpredicts the number of candidates by 15\% overall.
This discrepancy is beyond the standard uncertainty estimates for models
of neutrino interactions and neutron propagation in the detector.
We explore these two aspects of the models using the measured distributions
for energy deposition, time of flight, position, and speed.
We also use multiplicity distributions to evaluate the presence
of a two-nucleon knockout process.
These results provide critical new information toward a complete description of
the hadronic final state of neutrino interactions, which is vital
to neutrino oscillation experiments.
\end{abstract}
\ifnum\sizecheck=0
\maketitle
\fi

\section{introduction}

Neutrons are the last essential (anti)neutrino interaction final state particle to have their number and
energy distribution studied in detail. 
Neutrons play a special role in
those oscillation measurements which depend on comparing distortions
of the antineutrino energy spectrum to the neutrino energy
spectrum~\cite{ Abe:2017uxa, Adamson:2016xxw, Adamson:2016tbq, Acciari:2015zzz,
Adams:2013qkq,Acciarri:2015uup}.
The antineutrino reactions' calorimetric response is heavily
suppressed relative to the neutrino case because of the prevalence of
neutrons,  consistent with their generic role in hadron calorimetry \cite{Wigmans:2017vgs}.
The neutron energy deposits are not proportional to
their kinetic energy and are not always observed at a location consistent
with their initial trajectory from the interaction point.  Sometimes they are
not observed at all because they escape the detector volume unseen,
their interaction products are
below detection threshold, or they thermalize without producing
enough ionization to be reconstructed.
In addition to energy determination, neutrons' presence in the final state
also impacts event selection and background
rejection for oscillation, interaction, and rare process analyses.

In this paper, we present the first direct measurements of the neutron content
from neutron-rich antineutrino charged-current reactions.
Neutrons with ten to hundreds of MeV
kinetic energy are observed  as they
rescatter off hydrogen and carbon in the MINERvA detector.
This study is done with a low
three-momentum transfer sample \cite{Gran:2018fxa,Rodrigues:2015hik} 
but is otherwise inclusive with no selection on the number or type of
hadrons in the final-state.   
Low overall hadronic activity means that neutron activity is easily
separated and likely to be due to neutrons from the original
neutrino interaction rather than secondary hadronic reactions.
The results are compared to the predictions of a full simulation that
consists of a modified version of the {\small GENIE} neutrino
event generator \cite{Andreopoulos201087},
a G{\small EANT}4 simulation of particle propagation in the detector 
material \cite{Agostinelli2003250,1610988}, and a
simulation of the calibrated response of our scintillator and
electronics \cite{Aliaga:2013uqz,Aliaga:2015aqe}.

The energy deposit, time, position, speed, and multiplicity distributions are sensitive to the
details of neutron production in the initial reaction and to models for
neutron scattering in the detector.
When the neutrino reaction occurs in a nucleus, modeling the hadronic
final state is complicated.   
Different charged-current weak-interaction processes produce different
numbers of neutrons after the final-state lepton gains its charge.
The charge-changing
antineutrino quasielastic (QE) and two-particle two-hole
({\it 2p2h}) process must turn at least one proton into a neutron.
Often the neutron has
all of the hadronic final-state energy. 
In resonance production and deeply inelastic scattering (DIS)
the charge can be exchanged with the resulting meson or the struck
nucleon or quark,  producing a higher number
of neutrons per event than the equivalent neutrino case.
These outcomes for carbon are summarized in Table~\ref{tab:neutroncounts}.
For antineutrino reactions on hydrogen, the {\it 2p2h} process and final-state
interactions (FSI)
do not occur; also charged-current neutrino-hydrogen QE reactions are not possible.

\begin{table}[h]
\begin{tabular}{ccc}
  & \multicolumn{2}{c}{neutron content} \\
Process & Antineutrino & Neutrino \\ \hline
QE & 1 & 0 \\
Resonance & 1 or 0 & 0 or 1 \\
{\it 2p2h} & 2 or 1 & 1 or 0 \\
After FSI & 0 to 7 & 0 to 5
\end{tabular}
\caption{Typical number of free neutrons from
  charged-current antineutrino reactions with carbon compared to
  neutrino reactions.  In the extreme, FSI can break up the nucleus releasing all the
  neutrons. 
\label{tab:neutroncounts}}
\end{table}

Sensitivity to the mix of reaction types is reduced and indirect
because the resulting
hadrons will frequently reinteract on their way out of the nucleus.
In event generators, these rescattering processes are referred to as
final-state interactions (FSI).
Such reinteractions can be soft scatters that do not change the
outgoing charge state, full charge exchange reactions where the
energetic neutron becomes a proton or vice versa, and knockout
reactions where multiple nucleons and mesons exit the nucleus.
Calorimetric measurement of the interaction is affected and the hadron
topology also changes.

Neutrons from neutrino and antineutrino reactions have been measured
before.   The earliest technique was tagging the capture of
thermal and ``fast'' (up to 10 MeV) neutrons, used from the very
first neutrino observations
\cite{Cowan:1992xc} to the present and 
near future \cite{Amsbaugh:2007ke,Zhang:2013tua,Back:2017kfo,
  Beacom:2003nk}.
Higher-energy neutrons from neutrino reactions caused the most important
$np \rightarrow np$ background to the discovery \cite{Hasert:1973ff,Hasert:1974ju} of
weak neutral-current reactions in the freon-filled Gargamelle detector at
CERN.   The collaboration made measurements of ``associated neutrons'' and used a
cascade simulation \cite{Fry:1975qj} to translate that measurement into an
estimate for neutron production from neutrinos interacting in the
material upstream, and so obtained the crucial constraint on the
background.  
Similar studies were needed for
follow-up neutral-current measurements including
those with two liquid scintillator detectors
\cite{Entenberg:1979wc,Horstkotte:1981ne,Ahrens:1986xe} 
in the Brookhaven National Laboratory
neutrino beam, and the ``dirt backgrounds'' from MiniBooNE's
measurement \cite{AguilarArevalo:2010cx,Perevalov:2009zz}. 
These neutrons' time and spatial distributions in the detector were
measured and simulated in order to constrain and subtract backgrounds, but were
not correlated with simultaneous measurement of their original
interaction in the material upstream of the detector.
MiniBooNE's paper presents a comparison to a modern Monte Carlo
simulation using the {\small NUANCE} neutrino event generator
\cite{Casper:2002sd}, 
and G{\small EANT3} \cite{Brun:1987ma} using the {\small GCALOR} \cite{ZEITNITZ1994106}
option, and found that a 30\% reduction of the neutron component
was needed to describe the data.

Recently, a measurement \cite{Acciarri:2018myr}
was presented by ArgoNeuT
for neutrino-argon reactions which is similar to the one in this paper.
The ArgoNeuT analysis studied low-energy photons
produced by the deexcitation of the argon nucleus struck by the
neutrino and by the deexcitation of argon nuclei struck by neutrons
generated in the neutrino reaction.  Unlike the MINERvA antineutrino
data described in this paper, the neutron component is only half the sample, and the
photons from deexcitation of the argon nucleus struck by the neutrino
are evident in the spatial and multiplicity distributions.
In the ArgoNeuT study, both components were simulated by
{\small FLUKA} \cite{Bohlen:2014buj,Ferrari:2005zk}.
Another recent paper \cite{Friedland:2018vry} breaks down the
simulation of neutrino and antineutrino reactions with argon to
provide details of the pathways to missing energy, especially via neutrons.

MINERvA has several advantages relevant to detecting
neutrons from (anti)neutrino interactions.  The 5.3 ton, fully active
tracking volume is much larger than the neutron-interaction length
of approximately 10~cm
for neutron kinetic energies near 20 MeV.
Larger volumes have been used (Super-K and
MiniBooNE) but their Cerenkov technique has too high a threshold for
detecting the protons scattered by neutrons.  MINERvA's active volume
is polystyrene $(C_8H_8)_n$, shortened to CH when used in nuclear and
particle physics.   For low-energy neutron
detection, the hydrogen presents as
significant a target as the carbon, despite a carbon nucleus having 12
nucleons.  
And MINERvA is sensitive down
to a low threshold of 1 MeV because it is an underground detector
with low noise overall and well-constrained contributions from other
sources of neutron
production in the beam.

\section{MINERvA experiment}

MINERvA is a dedicated neutrino-nucleus cross section experiment.  Its
goals are to make cross section measurements needed for neutrino
oscillation experiments and to probe the environment of the nucleus, 
complementary to what the electron scattering community has
accomplished.  The experiment is located in the high-intensity
``Neutrinos at the Main Injector''
(NuMI) beam at Fermilab.  

The centerpiece of
the detector \cite{Aliaga:2013uqz} is a 5.3 ton fully active scintillator tracker with excellent
calorimetric properties of its own and surrounded by additional
electromagnetic and hadronic calorimeters. 
The experiment also has passive targets made of iron, lead, water,
graphite, and helium, which enable the study of the A-dependence of
neutrino and antineutrino reactions.

The tracker is built from planes of polystyrene scintillator strips.
With Lexan sheets, epoxy, tape, and reflective titanium dioxide, the target consists of
8.2\%, 88.5\%, and 2.5\% hydrogen, carbon, and oxygen respectively (by mass),
plus small amounts of heavier nuclei.
The strips are triangular in shape with 3.3 cm base and 1.7 cm height and up to
245 cm length. 
The strips are nested with alternating orientation to make
1.7 cm thick planes.   With this arrangement, ionization activity in
the plane is split between strips and tracking resolution is improved.
An entire plane is a hexagon containing 127 strips and one
module consists of two planes.  The second plane in one module is oriented with the strips vertical,
producing an X-coordinate of the observed energy deposits.  The
plane in front of it is rotated 60 deg one way to form a U coordinate or the
other way to form a V coordinate.   The resulting modules are
themselves alternated to produce repeating UX,VX sets of planes,
with the detector Z axis running normal to these planes.
The MINOS Near Detector \cite{Michael:2008bc} is located 2~m
downstream of MINERvA and measures the charge sign and momentum of muons
selected in this analysis.


These data were obtained from the NuMI
beam~\cite{Adamson:2015dkw} operating in antineutrino mode.
The primary 120 GeV proton beam interacts in a graphite target
producing pions and kaons.   Two magnetic horns focus the the
negatively charged mesons toward a decay pipe, leading to an
antineutrino spectrum that peaks near 3.0 GeV.   In total, these data
are from an exposure of
$1.04\times10^{20}$ protons on target between November 2010 and
February 2011.  This beam configuration also produces neutrino charged-current interactions in
the detector which are over 10\% of the events in the antineutrino
peak energy range used in this analysis.   These are
removed with high efficiency because their measured curvature in the
MINOS Near Detector is the wrong direction.

The flux prediction is
{\small GEANT4} based \cite{Agostinelli2003250,1610988} 
with central values and
uncertainties adjusted~\cite{Aliaga:2016oaz} using thin-target
hadron production data~\cite{Alt:2006fr, Denisov:1973zv,
  Carroll:1978hc,Allaby:1969de} and an {\it in situ} neutrino-electron
scattering constraint~\cite{Park:2015eqa}.  The design of this
analysis is relatively insensitive to the resulting 8 to 10\% uncertainties in the energy spectrum or
the absolute flux.

\section{Simulation}

The reference simulation combines the {\small GENIE} 2.8.4 neutrino-interaction
model \cite{Andreopoulos201087} with modifications, 
the {\small GEANT4} 9.4.p2 particle transport model
\cite{Agostinelli2003250,1610988} with modifications, a
calibrated detector time and energy response model \cite{Aliaga:2013uqz,Aliaga:2015aqe},
and an ``overlay'' of data events to reproduce
the unrelated activity that might overlap  in time with the simulated event.

\subsection{GENIE event generator}

{\small GENIE}'s simulation of the charged-current quasielastic
process is from Llewellyn Smith \cite{LlewellynSmith:1971zm} with vector form factors
parametrized as in Ref.~\cite{Bradford:2006yz}, and a dipole form factor with
axial mass of 0.99 GeV.   A relativistic Fermi gas model
\cite{Smith:1972xh} 
is implemented for interactions on carbon and other nuclei.
The $\Delta$ and
higher resonances are from Rein and Sehgal \cite{Rein:1980wg}, with a nonresonant
component added from the DIS model
as the resonances are phased out from invariant
mass 1.4 $<$ W $<$ 2.0 GeV.  The DIS
cross section comes from the Bodek-Yang model \cite{Bodek:2004pc},
where the hadronic system \cite{Yang:2009zx}
is produced using Koba, Nielsen, and Olesen (KNO) scaling \cite{Koba:1972ng} transitioning to
{\small PYTHIA} \cite{Sjostrand:2006za}
between 2.4 and 3.0 GeV. 

Modifications are
made to the above default {\small GENIE} 2.8.4 model.
We refer to this set as MINERvA tune version 1.1 ({\small MnvGENIE}-v1.1). 
These modifications can also be applied to the default 2.12.6 version
of {\small GENIE}.  
The QE process is modified to include a screening effect based on
the random phase approximation (RPA) technique.  The suppression is 
based on the calculations of Nieves and collaborators
\cite{Nieves:2004wx,Gran:2013kda} for a Fermi-gas nucleus and
implemented by reweighting {\small GENIE} QE
events \cite{Gran:2017psn}, including an uncertainty on the RPA screening
derived from comparison to neutron capture data \cite{Valverde:2006zn,Nieves:2017lij}.
The uncertainty on the QE axial form factor is set to 9\% following the
analysis of Ref.~\cite{Meyer:2016oeg}; additional uncertainty on the
highest $Q^2$ component is not needed for this sample.
Nonresonance pion production is reduced based on the reanalysis of
bubble chamber neutrino data \cite{Rodrigues:2016xjj,Wilkinson:2014yfa}.
Coherent pion production with pion kinetic energy below 450 MeV is
also reduced based on analysis of MINERvA
data~\cite{Higuera:2014azj,Mislivec:2017qfz} and
consistent with the Berger-Sehgal \cite{Berger:2008xs} modifications of the original
Rein-Sehgal model \cite{Rein:1982pf}.

Of special interest is the interaction of
the antineutrino with two nucleons, 
knocking them both out and leaving two holes
({\it 2p2h}) in the nucleus.
In this analysis, the base model for the {\it 2p2h} component with no pion is from 
the IFIC Valencia group \cite{Nieves:2011pp,Gran:2013kda}
implemented in {\small GENIE} \cite{Schwehr:2016pvn}.  This process is further
enhanced in the region between the QE and $\Delta$ resonance
components based on a fit \cite{minerva2p2h:2019} to the reconstructed
neutrino data presented in Ref.~\cite{Rodrigues:2015hik}.
In all but the last figure in this paper, the error band includes an
uncertainty on this
fit that varies the fraction of reactions on {\it pn} and {\it pp} initial states.
This enhancement has been applied to this analysis and successfully
describes a wide variety of data for other MINERvA observables \cite{Betancourt:2017uso,
  Altinok:2017xua, Patrick:2018gvi, Lu:2018stk, Gran:2018fxa, Ruterbories:2018gub}.

\begin{figure}[tbh!]
\begin{center}
\includegraphics[width=8cm]{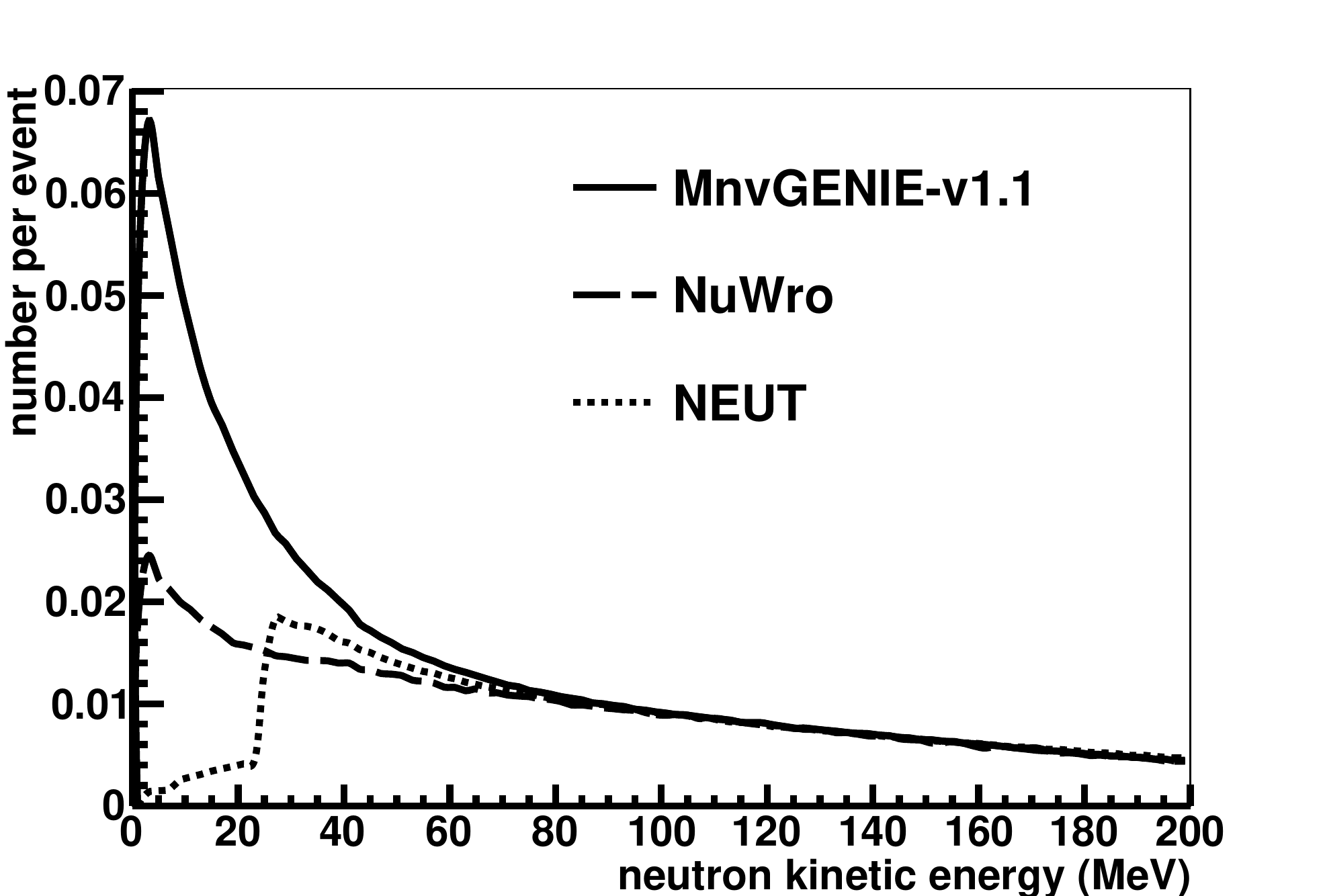}
\includegraphics[width=8cm]{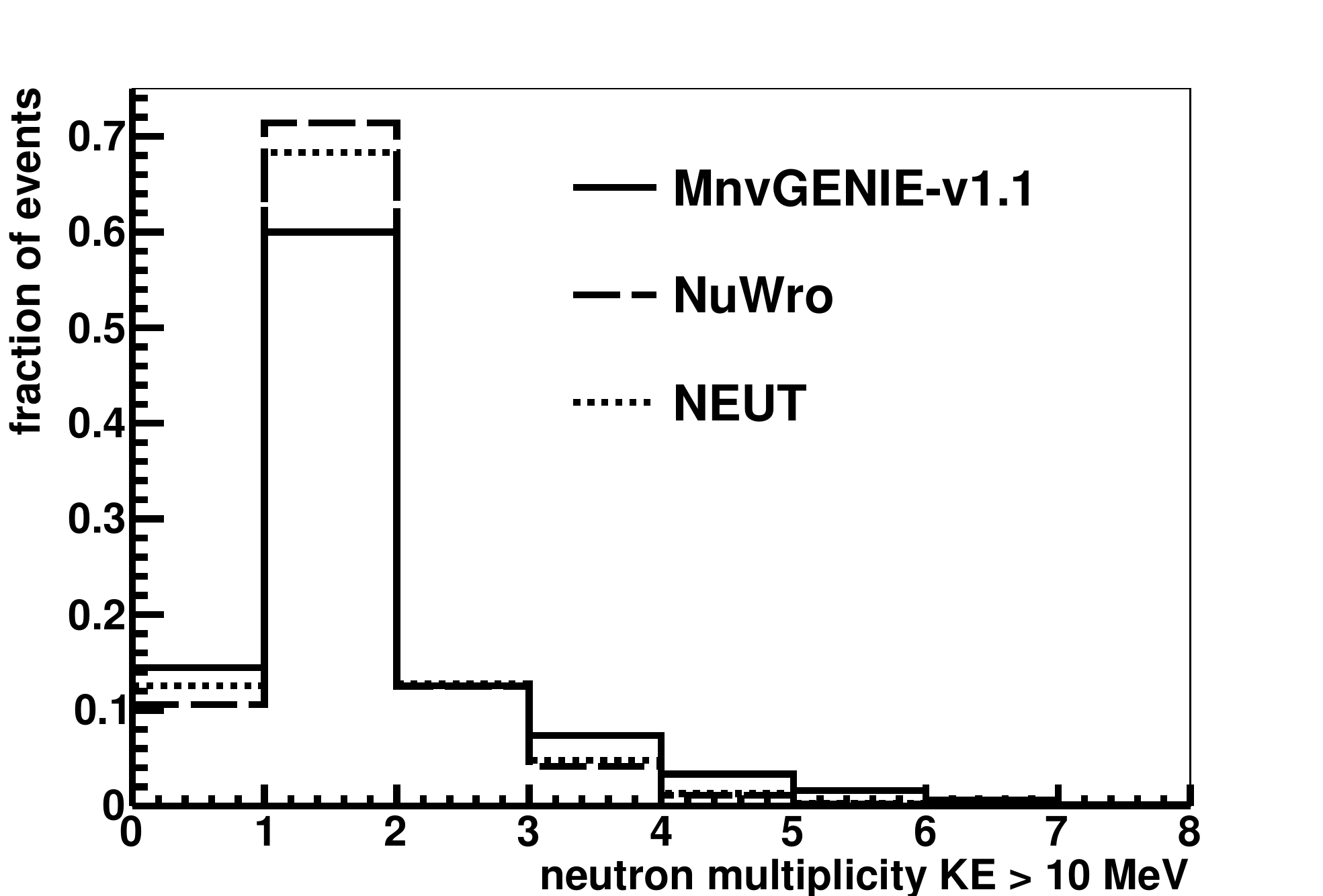}
\caption{Neutron energy spectra (upper) and multiplicity (lower) from three popular neutrino event
  generators for 3 GeV antineutrinos interacting in CH.   There
  exists a wide range of predictions, especially for
the lowest-energy component of this spectrum (see text for
description). 
\label{fig:truthmultiplicity}}
\end{center}
\end{figure}

Reinteractions of nucleons and mesons FSI  as they exit the 
target nucleus are the other important mechanism for neutron
production.
The {\small GENIE} simulation of final-state reinteractions of hadrons leaving
the nucleus is a  parametrized, effective cascade model which
is called ``hA'', short for hadron-nucleus interaction.
The model steps hadrons through a nucleus with its associated radius
and nuclear density function.   The hadron's mean free path is then
determined from tabulated hadron-proton and hadron-neutron cross
sections from SAID~\cite{SAID}.  The resulting probability of interacting
within the nucleus is high, 73\% for a neutron from a 3 GeV
quasielastic event in carbon, and 88\% in iron.   Because the code
originated with the MINOS experiment~\cite{Michael:2008bc}, when an
interaction is specified, the fates (absorption, pion production,
knockout, charge exchange, elastic scatter) are chosen according to
their proportions for iron.  In-medium effects that might relatively
favor particular fates in carbon or lead are not included in this
original version, nor is multiple scattering.
When one interaction occurs, the number of outgoing
nucleons is drawn from a distribution that favors single nucleon,
deuteron, and alpha states, but allows a chance for complete breakup.

Compared to the cross sections for the initial reactions, 
the simulation of FSI differs the most among different
event generators.  The {\small GENIE} FSI model produces
more low-energy nucleons than other commonly used neutrino event generators.
Figure~\ref{fig:truthmultiplicity} shows the energy and
multiplicity spectra for a newer (2.12.10) but equivalent version of {\small GENIE} with its models
and modifications configured for the {\small MnvGENIE-v1.1} tune.   It is
compared to two other common
generators {\small NEUT}~\cite{Hayato:2002sd,Hayato:2009zz} and
{\small NuWRO}~\cite{Golan:2012wx},  produced using the
{\small NUISANCE} framework~\cite{Stowell:2016jfr}. 
The inputs are monoenergetic 3 GeV antineutrinos
on a CH target, and the $q_3<0.8$~GeV selection is made.   Neutrons below 2
MeV are not included in the kinetic energy distribution.  Appropriate
to this analysis, neutrons below 10 MeV are not included in the
multiplicity distribution.
Below 25 MeV, the {\small NEUT} generator stops any reinteraction process
(cross section is set to zero), so
those neutrons are placed outside the nucleus rather than producing
yet more and lower-energy neutrons.
In contrast, above 50 MeV, each generator is using similar models for
the primary production of neutrons before FSI, and their predictions 
for the neutrons' kinetic energy spectra converge.

\subsection{G{\small EANT}4 neutron propagation model}

Particles in the final state are passed to G{\small EANT}4 9.4.p2 with the
Bertini cascade option ({\small QGSP\_BERT}) which
propagates them through a model of the MINERvA detector geometry and materials.
Uncertainties in the charged hadron-interaction cross section in the scintillator are implemented using a
reweighting scheme and an uncertainty of $\pm10$\% based on our
analysis of this version of {\small GEANT} and data for protons
\cite{Dietrich:2002, Menet:1971zz, McGill:1974zz,
  Renberg:1972jf,Dicello:1967zz, Zanelli:1981zz, MacGregor:1958zz, Bauhoff:1986gcb}
and
pions \cite{Ashery:1981tq, Wilkin:1973xd, Clough:1974qt,
  Allardyce:1973ce, Gelderloos:2000ds}.
Confidence in the proton and pion response at these
energies also comes
from analysis of calorimetric data taken with MINERvA detector elements in a test
beam at Fermilab \cite{Aliaga:2015aqe}.   
The charged hadron scattering cross sections in the detector are not
major sources of uncertainty.
The same study found the neutron inelastic scattering
on nuclei also models the data \cite{Ibaraki:2002,Schimmerling:1973bb,
  Voss:1956} at the 10\% level, but the total cross section (elastic +
inelastic) for this version does not describe the Abfalterer data \cite{Abfalterer:2001gw}.


The energy dependence of the total cross section for
neutrons in G{\small EANT}4 has been improved since this older version.
We have compared this version with a late 2016 release
(v10.2.p2) to understand changes in the neutron-interaction code.
The new versions now
match the Abfalterer {\it et al.} high-precision neutron scattering
total cross sections
\cite{Abfalterer:2001gw} on carbon and hydrogen from 5.2 to 560 MeV.   
Using the same reweighting
tools as to evaluate hadron systematic uncertainties, 
the older G{\small EANT}4 cross section is changed to represent these same data.  After
making this correction we assign uncertainties of 25\% below 10 MeV,
20\% from 10 MeV to 25 MeV, and 15\% above 25 MeV.    These 
are applied to the total cross section, but are larger than the
remaining discrepancy with the data.  In this
analysis they play the effective role of uncertainties on the elastic
vs. inelastic components.

In principle, another uncertainty arises from the outcomes of the
neutron-hydrogen and neutron-carbon interactions.   
The Abfalterer {\it et al.} total cross section includes elastic scatters that
deflect the neutron as little as 0.12 degrees, which involve sub-keV scale
energy transfers.  Thus, the G{\small EANT}4 model also makes a prediction for the
fraction of scatters that are above and below our experimental
threshold.   The accuracy of this feature of the prediction is not
well constrained by available data. 

A second, prominent feature of the G{\small EANT}4 neutron-scattering
model is the production of
deexcitation photons following neutron-carbon reactions.  Such
photons Compton scatter and account for 50\% of the
neutron candidates that originate from a {\small GENIE} neutron.  The
simulation predicts the rest are from protons and up to 5\% from
nuclear fragments.
For comparison, three recent accountings of neutron induced activity are
presented in \cite{Acciarri:2018myr, Bhandari:2019rat, Friedland:2018vry} for argon; the
latter has especially detailed discussion of simulated activity.

G{\small EANT}4 provides an alternate, 
high precision neutron simulation (HP in the G{\small EANT}4 option names), 
originally designed for studies of fission reactions up to about 20
MeV.  After accounting for the Abfalterer cross sections, the
differences between the fully simulated HP configuration and the default
configuration appear where the lowest-energy neutrons are
substantial parts of the sample, especially near the interaction point.  But these
differences are modest, similar to the other uncertainties, and difficult to
disentangle from the rest of the G{\small EANT}4 predicted response.

\subsection{Detector response}

The simulation also must produce
energy deposits consistent with our calibrated scintillator and
electronics response.  The photoelectron yield and absolute energy scale are tuned using a
comparison of data and simulated muons at near-normal incidence to the
planes.  Nonlinearities for energy deposits above minimum ionizing
muons are accounted for via individual calibration
of the digitizing electronics \cite{Aliaga:2013uqz} and by the test beam calibration \cite{Aliaga:2015aqe}
of Birk's quenching in the scintillator.

The simulation reproduces the detailed response of light propagation
in the scintillator bars and the photomultiplier tube, so that the simulated activity is reconstructed
using identical steps as with the data.  When a particle is fully
tracked, every energy deposit's location is known in all three
dimensions.  In this case, the reconstruction estimates the effect of
light reflection and attenuation in the scintillator strip and optical
fiber and
produces a more accurate estimate of the actual energy deposit.  This
is rarely possible for the neutron energy deposits in this analysis.   When
not a part of the track, the reconstruction approximates each energy
deposit to happen at a position halfway along the scintillator bar.
Geometrical fluctuations will therefore be present in the energy and
timing distributions in both data and simulation.

The width of the time distribution of digitized activity after light
propagation depends on the photoelectron yield in the photomultiplier tube.   The
distribution used in the simulation is based on the observed
distribution in data from fully tracked muons.  Because averaging
over the entire track yields a precise time and location for the passage of the
muon through any one plane, the correlation of the fluctuations in time
and the light yield in a single scintillator strip can be obtained
directly, without resorting to 
individual photon modeling of the optical elements and photomultiplier tube response. 
The width of the time distribution is 10~ns for 1 to 2
photoelectrons and 3~ns for 6 to 12 (about 1 to 1.5~MeV in a single
strip),  then approaches the electronics limit of 2.2~ns.
More detail on the nanosecond timing response and its use
can be found in Ref.~\cite{Marshall:2016rrn}.  With geometry-based
fluctuations, the final time resolution for neutron candidates is 4.5~ns.

A subdominant contribution to neutron candidates in this analysis
comes from other neutrino interactions inside and outside the detector, which
produce their own neutrons and photons.  They randomly happen at the
same time as the charged-current reactions selected for this
analysis.   These ``accidental'' backgrounds are added directly
to the simulation; 16~$\mu s$ of
activity from one pulse of the beam from data are
selected randomly from the same months.
This data activity is added on top of the activity from the simulated event.
The
resulting set of reconstructed times, locations, and energies are
given to the same reconstruction algorithm.   Thus this 
background and its dependence on the intensity of the beam
are reproduced.   Using alternate selections to isolate four regions
that are predicted to be high (greater than 70\%) in this particular background
leads to the assignment of a $\pm$10\%  uncertainty applied to the simulation.



\subsection{Further modifications of the simulation}

The configurations within the simulation packages do not contain enough explicit
uncertainties or tunable parameters to describe the data presented in
this paper.
Described in detail with the data in later sections, 
we will make two heuristic reductions in the number of
neutron candidates in the simulation.   One mimics a mismodeling of
either or both the energy deposit spectrum and the number of
deexcitation photons as simulated by G{\small EANT}4.   
The other reflects the wide range of predictions for low-energy
neutrons from the neutrino
interaction models
suggested by Fig.~\ref{fig:truthmultiplicity}.
For this paper,  the reductions allow us to
quantitatively explore possible unknown effects, even though they do
not represent a $\pm1\sigma$ uncertainty.

\section{Event sample and neutron selection}

The sample of charged-current antineutrino events analyzed in this
paper is the same inclusive
sample from Ref.~\cite{Gran:2018fxa} with reconstructed three-momentum
transfer $q_3$ less
than 0.8 GeV/c.   This sample has high neutron content but little other
charged hadronic activity to complicate the analysis of neutron
activity.  This allows us to use a low threshold of 1.5 MeV for
neutron candidates.

\subsection{Selection of antineutrino events}

Charged-current antineutrino interactions originate in the scintillator in the
active tracker fiducial region.   The resulting $\mu^+$ must be fully
tracked to the end of the MINERvA detector and also
reconstructed in the MINOS Near Detector \cite{Michael:2008bc} 
where its positive charge and momentum are analyzed.  
To ensure a region of good geometrical acceptance we require 
$p_\mu > 1.5$ GeV/c, and $\theta_\mu < 25$ deg.  
The selected reconstructed neutrino energy range is 
$2.0 < E_\nu < 6.0$ GeV,  so resulting data and simulated samples are
 the lower panels of Fig. 3 of Ref.~\cite{Gran:2018fxa} and
also match the selection used for the
related neutrino-mode analysis \cite{Rodrigues:2015hik}.

The muon energy and angle are combined with
the observed hadronic energy deposits to form calorimetric estimates for the energy
transfer $q_0$ (often called $\omega$ or $\nu$ by different groups in the literature),
neutrino energy $E_\nu = q_0 + E_\mu$, 
square four-momentum transfer
$-q^2 = Q^2 =
2E_\nu(E_\mu - p_\mu\cos\theta_\mu) - M^2_\mu$,
and the magnitude of the
three-momentum transfer  $q_3 = \sqrt{Q^2 + q_0^2}$ (often simply
called $q$ or $|q|$).
The selection is inclusive because only the magnitude of
reconstructed $q_3 < 0.8$ GeV/c enters the selection, 
not details of the number or type of hadrons observed.  

We exploit the feature that this subsample can be
divided into regions, hereafter called QE rich, dip, and $\Delta$ rich.
The QE-rich region has little or no observed hadronic energy and is
predicted to be mostly {\it 2p2h} and QE.   The $\Delta$-rich region has the
most energy transfer and is predicted to be mostly resonance production and some
{\it 2p2h}.  The so-called dip region in between these two is a mix of
all three processes, but also has the highest predicted concentration of {\it 2p2h} events.
Again referring to the lower panels of Fig. 3 in Ref.~\cite{Gran:2018fxa}, boundaries are formed between the QE-rich, 
dip-region, and $\Delta$-rich subsamples.  
The reconstructed ``available energy'' is an estimator for a
quantity that includes proton and charged pion kinetic energy and the
total energy of neutral pions, photons, and electrons produced by a
neutrino-interaction model.   The latter, built from the model
prediction for each generated event, explicitly does not include kinetic
energy of neutrons nor the energy cost to remove nucleons from the
nucleus, so it is always lower than the true energy transfer, and for
some antineutrino reactions will be zero.
The boundaries are at
reconstructed available energy of 0.06 and 0.12 GeV for $0.0 < q_3 < 0.4$ GeV/c
and 0.08 and 0.18 GeV for $0.4 < q_3 < 0.8$ GeV/c.
For brevity
many distributions in this paper are not divided this way and are
presented as 
two $q_3$ regions.

\subsection{Selection of neutral particle induced candidates}

Because the sample is limited to $q_3 < 0.8$ GeV/c, charged hadron
activity is small and remains localized to the interaction point.  The
rest of the detector should have no hadronic activity except neutrons and
photons, which this analysis considers signal.  There are two
backgrounds to neutron candidates to minimize: the ``muon background'' from electrons and
bremsstrahlung photons and the ``accidental background'' from activity 
induced by the neutrino beam or cosmic rays and unrelated to
the antineutrino reaction being analyzed. 

Analysis of hadronic
activity is done with
reconstructed objects formed of ``clusters'' of energy deposits in
one or multiple adjacent strips within the same plane.  Each cluster is assigned a
calibrated energy based on the sum of the individual energy deposits.
A cluster also has a two-dimensional position, one from the location
of the plane in the detector and one from estimating the
transverse position in its plane based on energy-weighted
average positions of activity in multiple strips.  The calibrated time is the
weighted average of hit times, taking into account the measured correlation between the
number of photoelectrons
and the width of the timing distribution of single strip activity from muons in data.  

\begin{figure}[tbh!]
\begin{center}
\includegraphics[width=8.5cm]{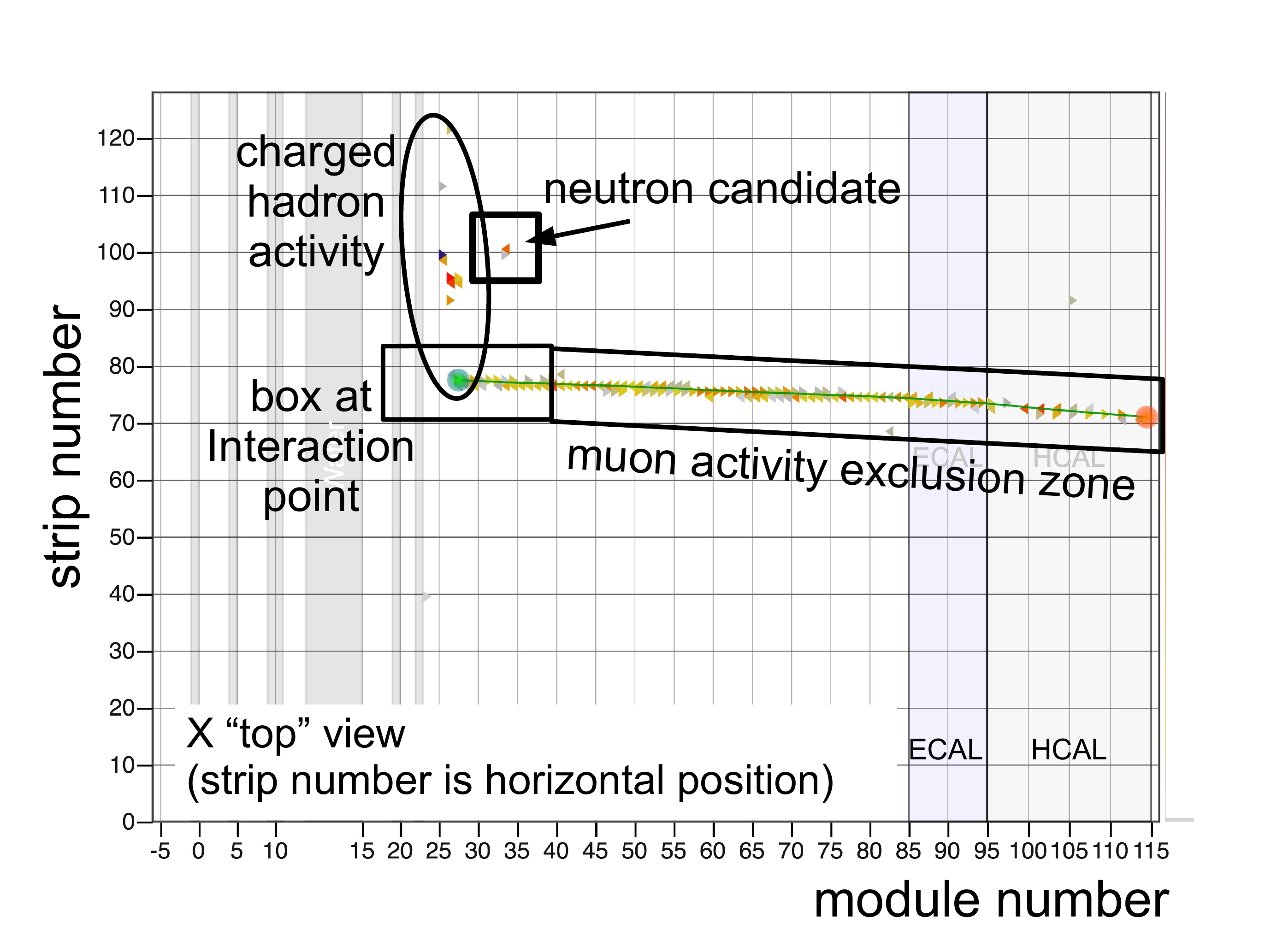}
\caption{Event display of a simulated event illustrating geometrical selections to
  avoid activity near the muon, event interaction point, and other charged
  hadron activity (a $\pi^-$ in this simulated event) with the
  remaining activity promoted to a neutron candidate.   The aspect
  ratio for this figure exaggerates the transverse dimension by 
  almost a factor of two in order to emphasize the detail.   Activity
  from the $\pi^-$ in adjacent U and V planes near the interaction point 
  is not shown.
\label{fig:eventdisplay}}
\end{center}
\end{figure}

The criteria for spatially isolated clusters are designed to exclude four overlapping
volumes in spatial proximity to the muon activity, 
the interaction point, other charged hadron activity, or to the
outer boundary of the detector. 
An event display of the X-coordinate activity simulated event,
Fig.~\ref{fig:eventdisplay}, 
summarizes how
activity in the first three categories is rejected and a single
cluster neutron candidate is observed.
Additional energy threshold and timing
cuts complete the selection.
Remaining clusters that are near each
other, indicating they may be caused by
the same neutral particle, are combined into multicluster candidates.
The simulation is used to evaluate the
effectiveness of these selections at reducing the muon and accidental
backgrounds.

\subsubsection{Muon exclusion zone}

The muon itself is fully tracked
for all events in this sample, and most clusters are already assigned
to the reconstructed track.  
Additional activity near the muon track is very likely caused by
photons from bremsstrahlung and knock-on electrons (delta rays).
From one module upstream of the interaction point to
the back of the detector 
we exclude clusters within 17 cm (about ten strips) of the muon from
consideration, about one mean free path for photons with energy of a few MeV.  
The simulation predicts that antimuon induced
activity accounts for 90\% of what would otherwise be candidates 
in this zone, but the simulation also underpredicts the data
by 12\%.
This exclusion reduces the muon-induced background by a factor of 10.

This zone is increased to 24 cm starting 20 modules downstream from
the interaction point, effectively a simplified
implementation of an exclusion cone.
The additional volume is predicted to be 65\% muon
activity and 3\% accidental backgrounds and excluding it further reduces
the muon background by another factor of 2.  The simulation describes the activity in
this outer zone well, contributing 3.5\% fewer clusters than data.  
We assign twice this difference as the uncertainty on the muon
contribution to the remaining selected clusters, which is negligible
for this analysis.

\subsubsection{Interaction point exclusion zone box}

Charged hadrons will produce activity near the neutrino-interaction
point and the start of the muon track.  The granularity of
the detector imposes limits on separating muon, hadron, photon, and
cross-talk activity when some or all of it fails to meet tracking
criteria.   This analysis follows our past
analysis strategy \cite{Fields:2013zhk,Fiorentini:2013ezn,Patrick:2018gvi,Ruterbories:2018gub}
to avoid this region and its complicated systematic uncertainties for a primary analysis.  The
neutrino-interaction model is most challenged when asked to predict
activity near the interaction point.   
This creates uncertain neutron detection efficiency effects and is deferred for future investigations.

Clusters of activity
within a transverse ``box'' around this point are excluded from further
consideration.   Because of the three X,U,V orientations, the box has
a pinwheel shape, but can be defined and coded simply when three-dimensional
reconstruction is not available. 
This exclusion is for clusters within 17 cm transverse
from a horizontal line parallel to the detector Z axis through the
start of the
muon track from ten modules upstream to fifteen modules downstream.
In the downstream direction, it usually overlaps significantly with
the muon exclusion zone.
This region remains rich
with neutron activity 
and may be explored in future neutron analyses.

\subsubsection{Charged particle exclusion zone}

Some charged hadrons travel outside the interaction point exclusion zone.   A spatial algorithm considers
seed clusters near the interaction point and adds additional clusters to the
total charged particle activity if they are nearby.   This algorithm
will follow such spatially connected activity arbitrarily far from the
interaction point.  
This includes many $\pi^-$ and a few protons that are fully tracked
but also activity from reinteractions in the detector that
do not satisfy stricter requirements to form or extend a track object.
This is the first full deployment of such an algorithm in a
MINERvA analysis.   

The procedure starts with a list of clusters that are inconsistent with cross-talk
based on their energy and pixel location relative
to other observed energy deposits within the same 64-pixel
photomultiplier tube.   Then, the two-dimensional
distance to the neutrino-interaction point is formed.   If at least one cluster is
found within 250 mm, the list is searched again and any cluster
within 100 mm of the first one is excluded from further
consideration.  Additional clusters are compared until no other cluster
satisfies the 100 mm requirement with any prior cluster.
Then it iterates five more times with the
original 250 mm cluster test, which may identify additional hadrons going off in
completely different directions.

For the many clusters near the interaction point, this procedure is largely redundant with the vertex
exclusion box.   However, it will follow spatially connected energy well outside the other exclusion zones.
This trivially includes energy that was already part of a hadron
track.  It also includes energy produced when a hadron reinteracts
creating additional untracked energy, and also hadrons from the
interaction point that did not satisfy the stricter requirements to
form a track object. 

The performance of this selection can be illustrated by the situations
that lead to a neutron candidate attributed to a $\pi^{-}$ from
{\small GENIE}, which is one
component of the signal and background that remain after all
selections and summarized in Table~\ref{tab:breakdown}.  After
inspecting the G{\small EANT}4 trajectory information, 85\% of those
neutron candidates from $\pi^{-}$ were caused by secondary neutrons, photons from
nuclear deexcitation, and
photons from the decay of neutral pions from charge exchange in the
detector.  The other 15\% were not incorporated into the charged
particle exclusion zone because some activity was highly transverse to
the detector.   There is such a simulated $\pi^{-}$ in the example in
Fig.~\ref{fig:eventdisplay}.  Unlike this example, there is sometimes a
cluster at high angle that did not meet the 100 mm tolerance to extend
the exclusion zone.  In total, these situations account for only 3\%
of the total candidate sample.


\subsubsection{Edge of detector, timing, and energy}

Activity at the edges of the detector is
excluded as follows:  the first 20 planes and veto wall upstream, the
last 10 planes in the downstream hadronic calorimeter, and the entire
outer detector hadronic calorimeter.    Most clusters are naturally
found in the inner tracker region because it is near to the
interaction point, has the largest fraction of the
active scintillator in the detector, and the scintillator has the
highest hydrogen content of all the detector materials.  


A few additional selections reduce the already small accidental
background.
Clusters not yet excluded must be within a time window from 20 ns before
to 35 ns after the interaction time $t_0$ determined largely from the muon
track timing information. 
The clusters must also have at least
1.5 MeV of energy.   The accidental background overtakes the predicted signal
processes at cluster energies below 1.2 MeV.  
This energy cut also eliminates photomultiplier tube
cross-talk effects in both
selected data and simulated events.
This version of the accidental background overlay technique is not perfect.  
For this early version it is checked against the data for accidental-rich
subsamples leading to an uncertainty of 10\%.
This conservative uncertainty has negligible impact on the results.

\subsubsection{Aggregating spatially nearby clusters into candidates}

Some isolated clusters can be spatially connected.
Two clusters within three modules
of each other are merged into one neutron candidate.  This merging continues
if additional clusters satisfy this requirement.
Because there is so
little hadronic activity in these low $q_3$ subsamples
this simple requirement is effective.
In this sample, 62\% are candidates made of a single cluster, 18\% are
made of exactly two clusters; the simulation predicts 61\% and 19\% respectively.

\subsection{Signal and background}

Neutrons that exit the nucleus where the neutrino interaction occurred are of
the most interest.  They are the aspect of the interaction model that
has never before been directly tested.  When referring to the {\small GENIE} simulation of this
component, including direct production and via FSI, we will call them {\small GENIE} neutrons. 
Protons and charged pions produce real secondary neutrons as they
travel through the detector.
These are an irreducible background 
and no attempt is made to reject them.   
Table~\ref{tab:breakdown} quantifies the sources that cause the
majority of neutron candidates.

\begin{table}[ht!]
\begin{tabular}{ccc}
 & \multicolumn{2}{c}{three-momentum transfer} \\
{\small GENIE} particle & $\;0<q_3<0.4\;$ & $\;0.4<q_3<0.8\;$ \\ \hline
Neutron & 78.1\% & 60.8\% \\
Proton & 0.4\% & 1.6\% \\
$\pi^-$ & 12.3\% & 22.8\% \\
$\pi^0$ & 2.6\% & 10.3\% \\
Muon & 4.0\% & 2.4\% \\
Data overlay & 2.1\% & 1.4\% \\
Other & 0.6\% & 0.6\% \\ \hline \hline
Events simulation &  4499 & 11651 \\
Events data & 4897 & 11263  \\
Cand/evt simulation & 0.647 & 1.284 \\
Cand/evt data & 0.584 & 1.103 
\end{tabular}
\caption{Particle from the {\small GENIE} simulation leaving the nucleus
 that caused the simulated candidates, showing the
  characteristics of the signal and background.  Candidates attributed
  to proton and $\pi^-$ from {\small GENIE} are largely caused by secondary
  neutrons.  The lower part of the table shows the number of selected events
  and neutron candidates per event (cand/evt) for this data and simulated exposure.  Compared to the data, the simulation has an overprediction
  of candidates caused by neutrons.
\label{tab:breakdown}}
\end{table}

Photons also produce energy deposits distant from the main parts of
the event.  Thus the decay products of $\pi^0$ 
are an electromagnetic component of candidates.   
The $\pi^-$ entries in Table~\ref{tab:breakdown}
include 14\% (so 1.7 and 3.2\% of the respective column totals) 
which charge exchanged to $\pi^0$ in the detector.
With such little hadronic energy in the final state, it is convenient to
treat these also as irreducible backgrounds, 
and make no selection to reduce them.  Because of the $q_3$ selection,
these photons are relatively low energy.
MINERvA's $\gamma$ and $\pi^0$ 
selection algorithms
\cite{Altinok:2017xua,Aliaga:2015wva}
have an efficiency of 15\% because $\pi^0$ identification
requires two reconstructed photon candidates and an effective
threshold of 25 MeV on the lower-energy photon.
As will be seen, the electromagnetic component traveling at the speed of
light will separate from slower neutrons in distributions that use
time-of-flight information.
Because no attempt is made to reject 
the $\pi^0$ backgrounds,
the analysis of neutral particle activity remains inclusive of all
hadronic final states.

The small ``other'' category originates with
$\pi^+$ and kaons, mostly from DIS processes which
feed down into the sample from higher $q_3$ reactions.   This 
category also includes photons and electrons from $\eta^0$ production and
$\Delta \rightarrow N \gamma$ decay.   
Photons from
{\small GENIE}'s deexcitation of residual nuclei would be present and isotropic in the data
sample, but are only simulated for oxygen, not carbon nor other
nuclei.   Radiative processes from the lepton exiting the nucleus
might be collinear with the muon and are also not simulated in {\small GENIE}.

The backgrounds from the muon
and accidentals are small.  The effectiveness of
the cuts can be illustrated quantitatively, relative to the base
selection. 
Including the very ends of the detector
doubles the accidental background.  Not excluding the 17 to 24 cm outer volume around the muon
doubles the muon background.   Using an incident neutrino energy range
up to 20 GeV increases the statistics of the total data and simulated sample by 17\% and
also all the simulated signal and irreducible background subcomponents.
Compared to the base selection, these higher-energy reactions have 30\% more in the
other category from feed-down of higher $q_3$ DIS events.  The muon
component (because bremsstrahlung is more likely) is also higher by
22\%.

Three other cuts could be relaxed to extend the physics reach, but doing
so would increase the backgrounds.   
These samples are consistent with the main sample
but do not further enhance the conclusions.
Reducing the energy threshold for candidates from 1.5 down to 1.2
MeV increases the predicted rate by 10\% overall, but the data rate
increases only by 8\%.   The predicted muon and
accidental background rate increase by 30\% and 50\% respectively, collectively accounting
for 20\% of the additional candidates.  Allowing
the timing to go out to 100 ns adds events that are predicted to be 
half from the accidental background.

\subsection{Efficiency}

\begin{figure}[tbh!]
\begin{center}
\includegraphics[width=8cm]{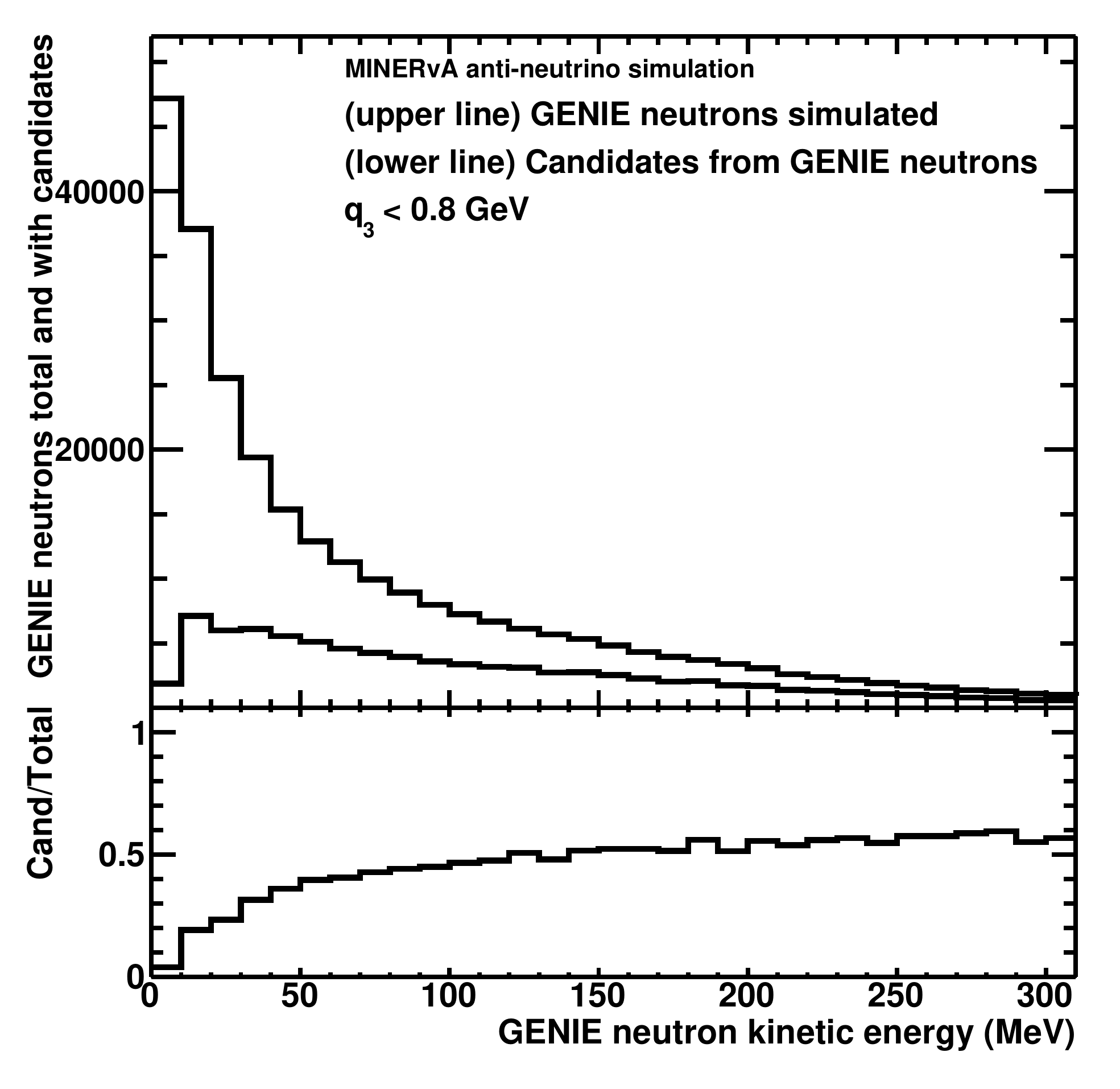}
\caption{Distribution of neutrons produced by the {\small GENIE} model for the
  selected $q_3<0.8$ GeV/c sample, and the subset of neutrons that produced one or
  more neutron candidates.  The ratio in the bottom panel is the
  efficiency to find at least one candidate from each neutron.
\label{fig:efficiency}}
\end{center}
\end{figure}

Focusing again on the most interesting signal process, the probability
that a neutron from the {\small GENIE} model produces a cluster of activity is
high because of the large volume of the fully active scintillator.
The probability to survive the selection
rises with kinetic energy from few to 60\% .   This is shown in
Fig.~\ref{fig:efficiency}, which also illustrates the predicted neutron kinetic
energy spectrum for this antineutrino $q_3<0.8$ GeV/c sample. 
Details of the selection process sculpt the distribution.
The region near the interaction point especially is a place where 45\% of neutrons leave
activity, as one would expect from a mean-free-path process.   Some lower-energy neutrons
are effectively below threshold after just one interaction, and the
lowest-energy neutrons do not travel very far and thermalize locally.
Higher-energy neutrons may still interact again and be visible.

The efficiency for detecting neutrons using plastic scintillator has
been studied since the late 1950s.  The 1962 measurement
\cite{WiegandNeutron} 
in a scintillator 15 cm thick with a simple discriminator-based
1.5 MeV electron-equivalent threshold is very similar to the conditions of
this MINERvA measurement.   They observed an efficiency of 20\% at 76
MeV and 30\% at 10 MeV.   Below this the threshold reduces the
efficiency significantly.   These data have been compared to
increasingly sophisticated simulations over the years, such as
in Ref.~\cite{Cecil:1979sz}.
G{\small EANT}4 produces a similar efficiency for neutrons above 20 MeV, but
predicts closer to 40\% efficiency at 10 MeV.

The mix of hydrogen and carbon in the detector affects the efficiency
for different ranges of neutron kinetic energy.
A special purpose G{\small EANT}4 neutron simulation in the MINERvA detector
with the same reconstruction as this analysis predicts that hydrogen produces
more candidates per nucleus up to neutron kinetic energy of 20 MeV where
it is equally probable as detectable scatters from carbon.   The
hydrogen-to-carbon ratio remains around 1:6 up to 100 MeV, then falls
further past 1:12 at higher neutron kinetic energies.
What is called neutron inelastic scattering in carbon begins around 10
MeV of neutron kinetic energy. (This includes single nucleon
knockout, which is the strong-interaction analog to the processes called quasielastic electron
scattering and neutral current elastic neutrino scattering.)   At lower
kinetic energy, neutrons are not able to transfer enough energy to a
proton to remove it from the nucleus, yet elastic scattering off
hydrogen can produce a proton above the detection threshold.
A liquid argon detector would have significantly less acceptance below
20 MeV kinetic energy for the same threshold; a liquid scintillator
(CH$_2$) detector could
have more.

\begin{figure}[tbh!]
\begin{center}
\includegraphics[width=9cm]{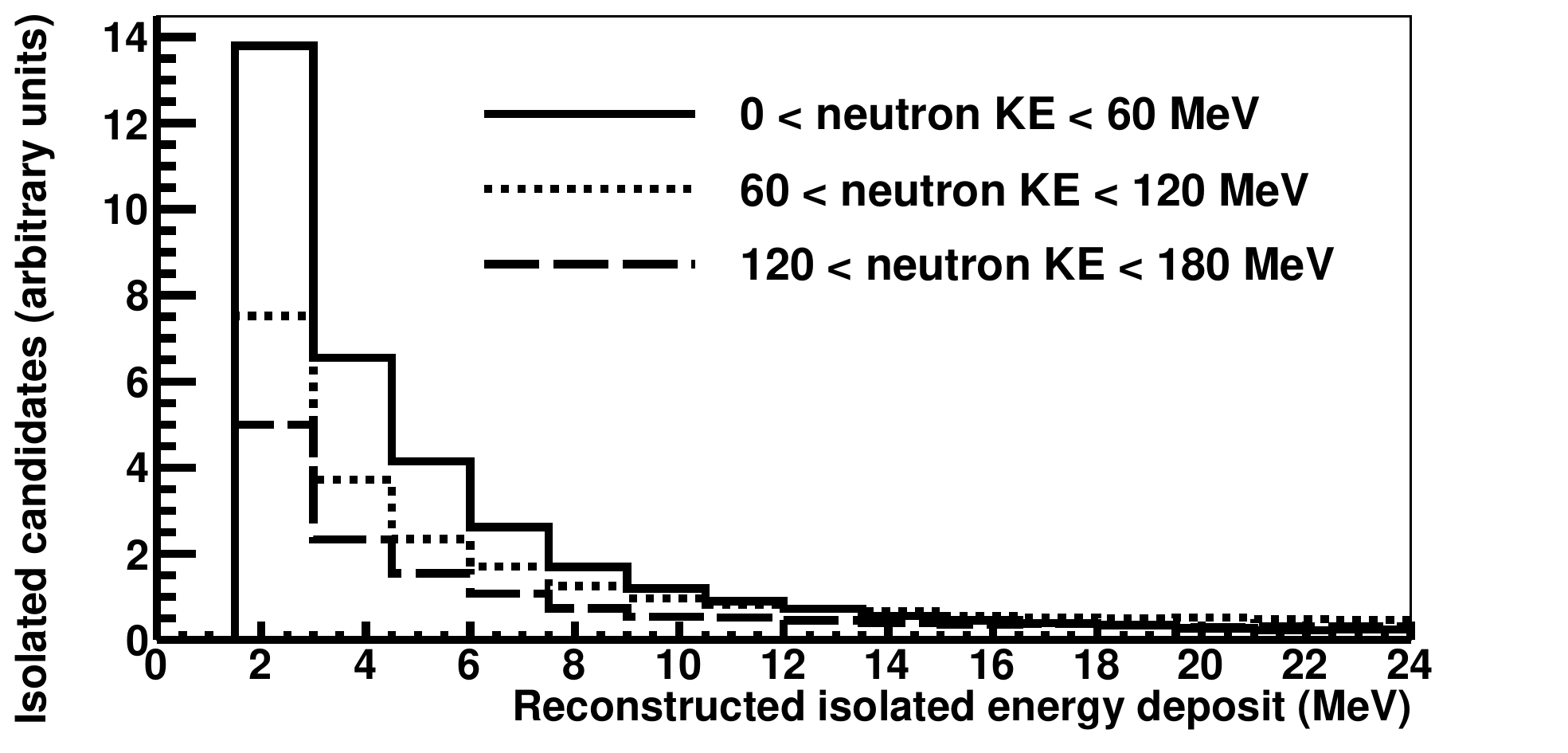}
\caption{Predicted candidate energy spectrum from {\small GENIE} neutrons in
  three energy ranges from the selected sample.  The similarity of the
  spectra prevent a robust, direct calorimetric neutron energy measurement.
\label{fig:energyVsKE}}
\end{center}
\end{figure}

Identifying the presence of a neutron is easier than guessing what the
energy of the neutron was.
The spectra of energy deposits for neutrons in three
different kinetic energy ranges are shown in Fig.~\ref{fig:energyVsKE}
(the same figure appears in Ref.~\cite{Gran:2018fxa}). 
The reconstructed energy of a single cluster, or the sum of two or more
clusters aggregated into a single candidate is mostly uncorrelated
with the kinetic energy of the neutron.
The most likely neutron candidate for all kinetic energies considered
in this sample is just at the 1.5 MeV threshold.

As the neutron energy rises, the probability it will produce an energetic
proton that travels several planes and leaves tens to hundreds of MeV
also rises, and populates the distribution at and beyond the right
edge of the plot up to hundreds of MeV.   

The converse is also relevant.  Neutrons in the range 10 to 20 MeV can
only make the smallest energy deposits.   The presence of a small
energy deposit does not automatically indicate a low-energy neutron.
But the presence of many low-energy neutrons can only enhance the
rate of the smallest energy deposits.

Overall, the presence of neutrons down to 50 MeV kinetic energy is
determined with good efficiency and low backgrounds. 
Though the efficiency continues
to fall, neutrons down to 10 MeV are predicted to cause a substantial part
of the sample.

\section{Measurements of neutron activity}

One observation already stands out from Table~\ref{tab:breakdown}.
There is an overall overprediction of neutron candidates in the simulation,
further emphasized in Table~\ref{tab:excess}.
It appears in all subdivisions of the sample, despite 
different amounts of QE, {\it 2p2h}, and $\Delta$ resonance, and their varying neutron
final-state content.
Because the efficiency is so high and the backgrounds are so low,
either {\small GENIE} is producing too many neutrons per
event or the
G{\small EANT}4 neutron propagation plus detector response simulation is making them more
visible than in the data.  
Compared to Table~\ref{tab:excess}, before modifying the G{\small EANT}4 cross section to match the
Abfalterer {\it et al.} measurements, the neutron candidates per event
were 1.20 for each range of momentum transfer.

\begin{table}[h]
\begin{tabular}{ccc}

 & \multicolumn{2}{c}{three-momentum transfer} \\
MC/data & $\;0<q_3<0.4\;$ & $\;0.4<q_3<0.8\;$ \\ \hline
Selected events & 0.92 & 1.03   \\
Neutron candidates & 1.02 & 1.20 \\
Neutron candidates/event & 1.11 &  1.16 \\
Statistical uncertainty & 0.02 & 0.01 \\
Systematic uncertainty & 0.07 & 0.04   
\end{tabular}
\caption{Ratio of simulation to data for selected events, neutron candidates, and candidates per
  event. 
All systematic uncertainties are accounted for except for two model
variations treated separately and described later.
\label{tab:excess}}
\end{table}


In this section, distributions of deposited energy, time, and position
upstream or downstream are shown.  Then 
neutron speed (actually 1/$\beta$) and multiplicity per event 
are used to draw final conclusions.   Starting with the time
distributions, the low-energy candidates and high-energy candidates
are separated, and the oversimulation persists in the former.

In all cases, the reconstructed data distribution is
shown with statistical uncertainties only (often too small to see) and
the simulation is shown with model and detector systematic
uncertainties.   Some physics effects being tested with these
data are not included in the systematic uncertainties and are
described explicitly.  Each figure is neutron candidates per event like
Table~\ref{tab:excess}, 
reducing systematics that affect the numerator and denominator
equally,  such as the flux and some cross section uncertainties.   
Discussion of specific uncertainties and resolutions
are provided with each new distribution, if not previously described.

In each plot, the simulation is broken down into {\small GENIE} neutrons,
other sources of neutrons, and the electromagnetic component.
The muon and accidental backgrounds are often too small to see and are
described in the text instead of being plotted.
Each figure has a lower, ratio subpanel where the data-to-simulation ratio is
compared to the reference model and its uncertainties, to emphasize the
magnitude and location of discrepancies.

\begin{figure*}[hbt!]
\begin{center}
\includegraphics[width=7cm]{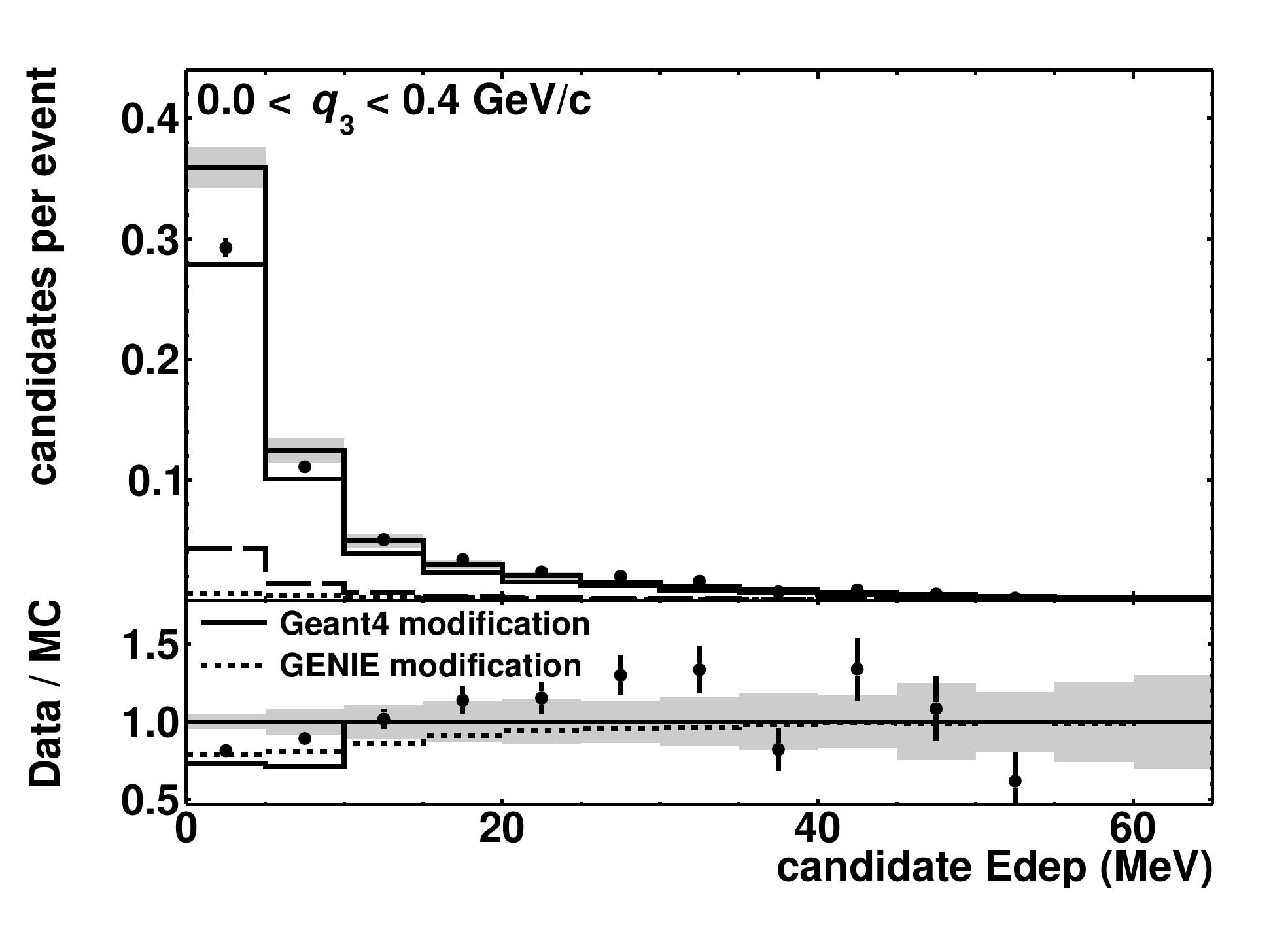}
\includegraphics[width=7cm]{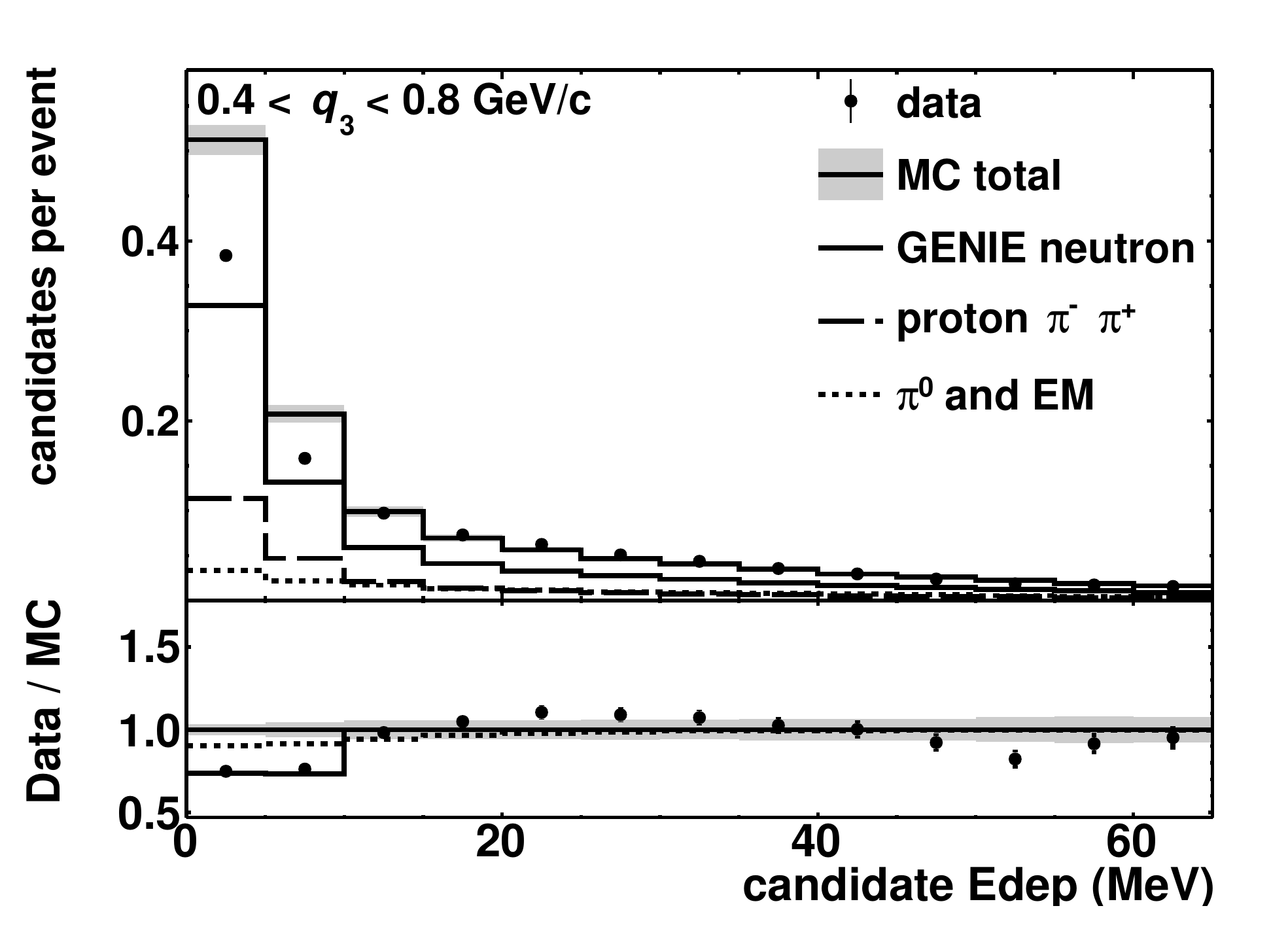}
\caption{Distribution of measured energy deposit $E_\mathrm{dep}$ per neutron
  candidate, normalized by the total
  number of events.  Data are shown with statistical uncertainties only; the simulation with is shown with systematic uncertainties. The lowest bin only contains candidates down to
  the 1.5 MeV threshold.
  The lower panels contain ratios to the reference 
  simulation for the data and for modifications to the {\small GENIE} and
  G{\small EANT}4 simulations which will serve as a benchmark
  and are described in the text and used hereafter.  Bins with very large data statistical
  uncertainties are not shown.
\label{fig:Energy}}
\end{center}
\end{figure*}

\subsection{Spectrum of candidate deposited energy}

The energy deposit spectrum $E_\mathrm{dep}$ shown in
Fig.~\ref{fig:Energy} highlights that the extra neutron candidates in
the simulation are limited to those with less than 10 MeV. 
They are unambiguously from neutron
production.  The predicted neutron components of the spectrum are much
higher and peak much more strongly at 
threshold than the electromagnetic component, shown by the dotted line
in Fig.~\ref{fig:Energy}.
The estimate of the backgrounds is not subtracted 
and is a 12$\pm1$\% contribution to the first
bin in each histogram (not shown with its own line, but similar to the
proton+pion population in those two bins), and negligible
everywhere else.  Above 10 MeV, the data are described by the
simulation, allowing for the systematic uncertainties summarized by
the shaded region.

We use these data to probe for model features beyond the standard systematic
uncertainties included in the error band.  The ratio subpanel includes two
modified models we will refer to as ``benchmarks'' throughout this section.  
For the solid ``modified G{\small EANT}4'' line, we eliminate a random 50\% of neutron candidates
with less than 10 MeV energy deposit and originating from
neutron-producing {\small GENIE} particles (neutrons, protons, $\pi^\pm$).
Without regard
to the energy of the particle, this mimics moving part of the G{\small EANT}4
or detector response below our detection threshold, making these elastic
scatters invisible, or reducing the number of photons produced by
nuclear deexcitations in carbon following the nucleon knockout process.   
The dotted line implements a neutrino-interaction model change;
50\% of candidates caused by {\small GENIE} neutrons below kinetic energy of 50
MeV are removed.  This makes the prediction more like NuWRO in
Fig.~\ref{fig:truthmultiplicity}.    
These benchmark modifications are chosen empirically to
have about the right size and allow the analysis to track their effects across the
rest of the distributions.  

Considering Fig.~\ref{fig:energyVsKE}, neutron-caused candidates above 10 MeV are  
necessarily from higher kinetic energy neutrons, while the lowest-energy  
candidates are a mix of everything.    
By itself, the modified {\small GENIE} benchmark that reduces only the
lowest-energy neutrons would provide a description the $0.0 < q_3 < 0.4$ GeV/c region at the edge of the systematic error band.   But
its effects are not strong enough to describe the right panel.  
The G{\small EANT}4 benchmark that removes only low-energy candidates
would describe both regions adequately,
and the 5 to 10 MeV point in the left plot would remain slightly outside the
error band.


\subsection*{Uncertainties}

Signal response
uncertainties that affect the probability a neutron will produce a
cluster near the 1.5 MeV threshold are important.
Neutron elastic scatters produce a low-energy proton of which the detector
response is affected by a scintillator quenching effect parametrized
using Birks's Law \cite{Birks:1951,Birks:1964zz}.  Our parameter and its uncertainties 
are calibrated using test beam data from stopping protons
\cite{Aliaga:2015aqe} especially the last 40 MeV of their energy deposits.   
The uncertainty is doubled for neutron-induced candidates with less than 5 MeV, which are not well
constrained by the test beam data.  
The total Birks' uncertainty changes the response as much as 25\%
for candidates near threshold.
Increasing the Birks suppression migrates
events down in these distributions and some fall below the 1.5 MeV
threshold.   Modern studies of Birks's quenching in liquid and plastic
scintillators \cite{vonKrosigk:2013sa, vonKrosigk:2015aaa}, made for
supernova and solar neutrino detection, are also at these energies and confirm
the predicted scintillator response.
This uncertainty is the largest single
contribution in the first bin of Fig.~\ref{fig:Energy}, but is only
4\%.

The G{\small EANT}4 cross section model uncertainties play a lesser role in
this distribution than they do in the time and spatial distributions
later.    An increase in the cross section
(decrease in the mean free path) makes candidates interact earlier.  Candidates are more likely in and near the
interaction exclusion region and there are fewer candidates overall.
It contributes 4\% to the uncertainty but only 2\% in the first bin.
The G{\small EANT}4 energy deposit model and photon yield are further 
explored using the benchmark modification shown with
the solid black line in the ratio panels.

Large rate uncertainties on the QE, {\it 2p2h}, and resonance models
combine for 3\% uncertainty on this distribution, but less than 1\%
uncertainty in the first two bins, and are the largest source for most bins
for $0.4 < q_3 < 0.8$ GeV/c.  All these antineutrino processes produce some neutrons
with similar energies.
Distortions of the
energy transfer spectrum for these processes,  such as the 
uncertainty assigned to the RPA screening effect for QE used in the reference model
\cite{Nieves:2004wx,Gran:2017psn} are at 4\% and only important
for the $q_3 < 0.4$ GeV/c subsample beyond the first two bins.

Effects related to the hadronic energy scale 
act in a special way and are significant for
energy deposits above 10 MeV.    They cause a
migration of events up and down the range of $q_3$.   
This migration effect is most significant and dominates parts of the error band
for $q_3 < 0.4$.  The subsample is lower statistics, has fewer candidates per event
overall, and does not have compensating event feed-in/-out at its lower boundary.
The hadronic energy scale
uncertainties are assigned based on test beam data with an enhanced 
uncertainty for the neutron response that was not directly tested.
The {\small GENIE} FSI
uncertainties also play this same migration role, in addition to directly changing
the number of candidates in each event, and contribute a similar amount
to total uncertainty.

\begin{figure*}[t]
\begin{center}
\includegraphics[trim=0 15 0 0,clip,width=7cm]{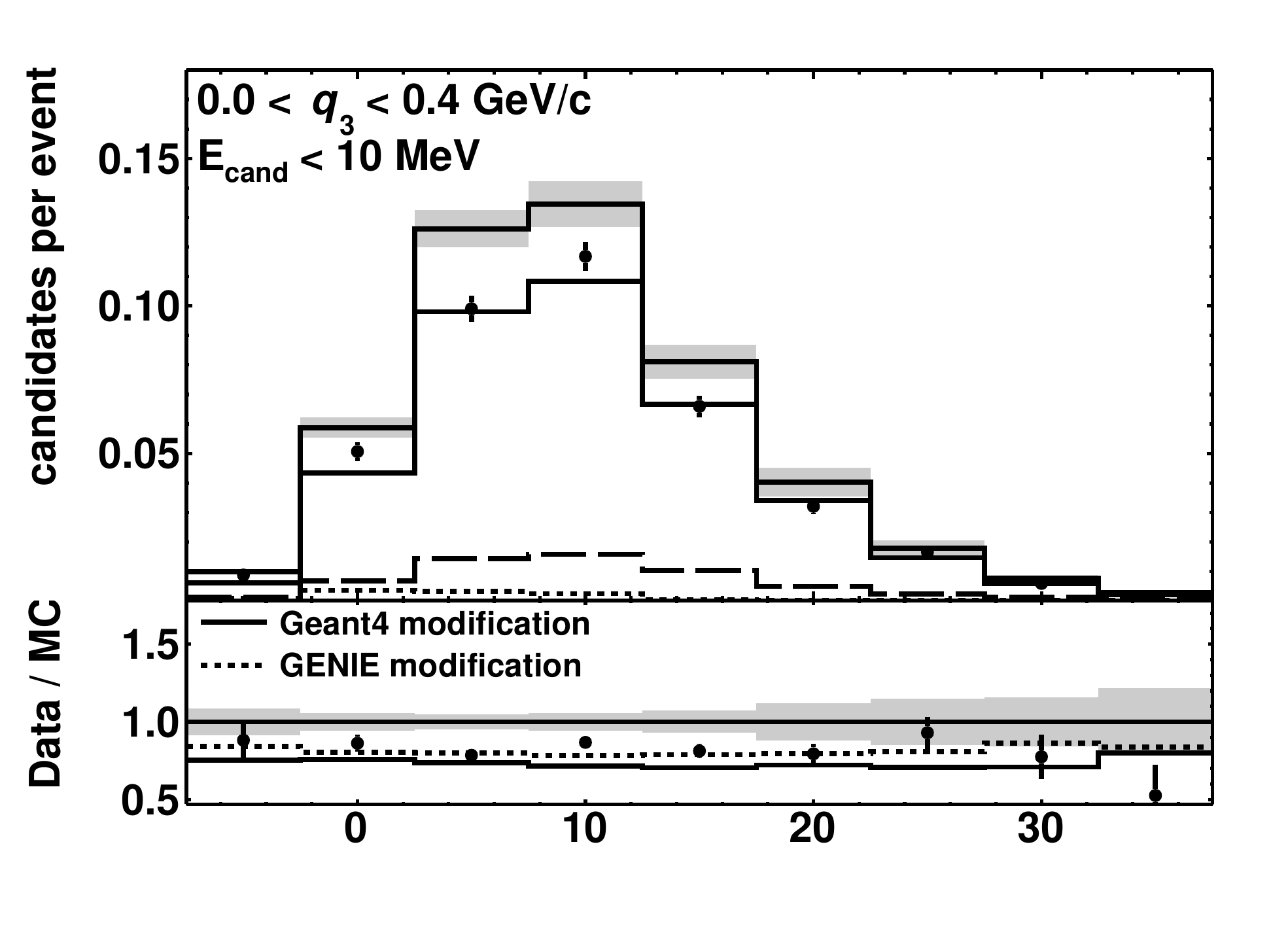}
\includegraphics[trim=0 15 0 0,clip,width=7cm]{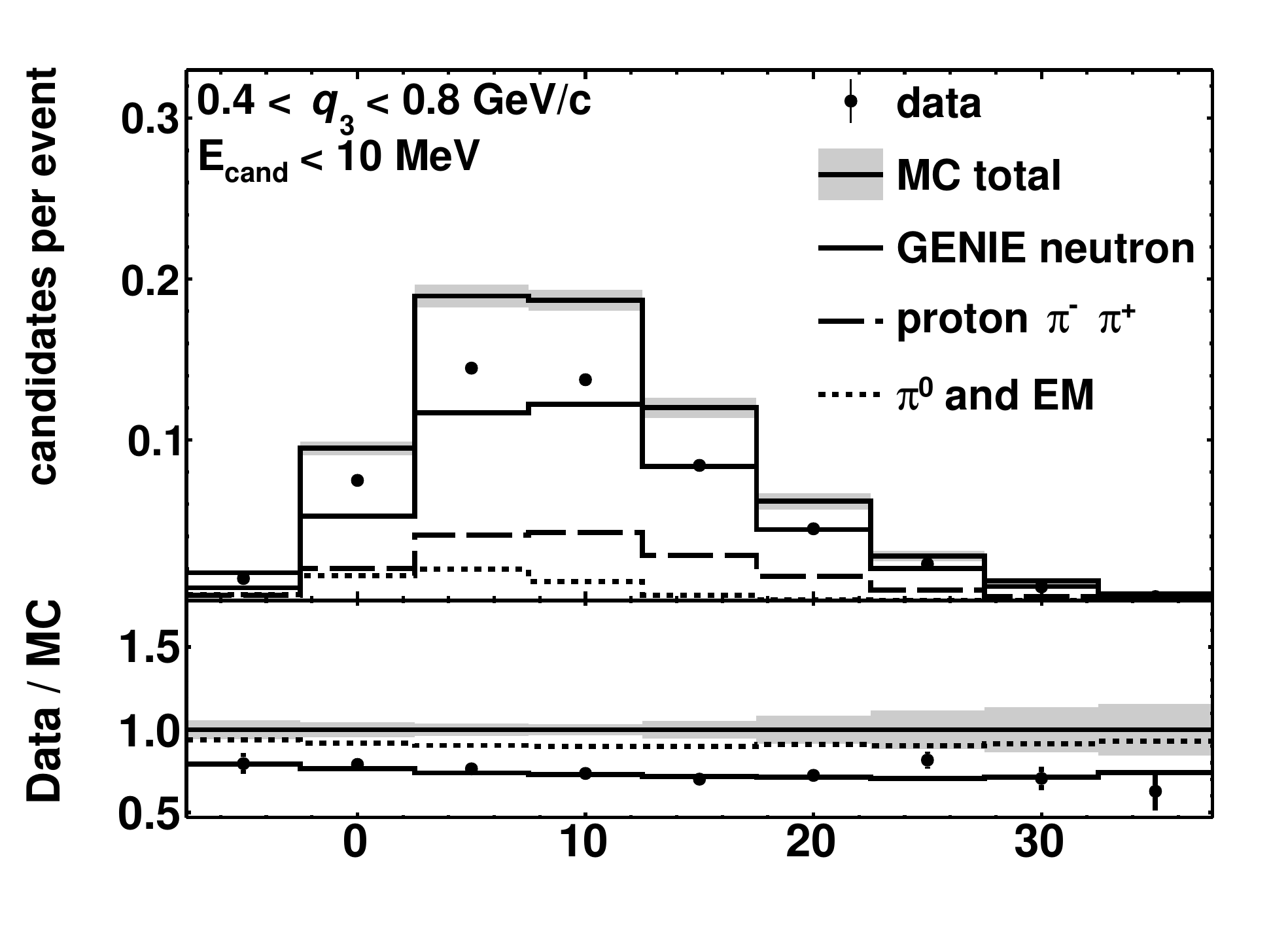}
\includegraphics[trim=0 0 0 30,clip,width=7cm]{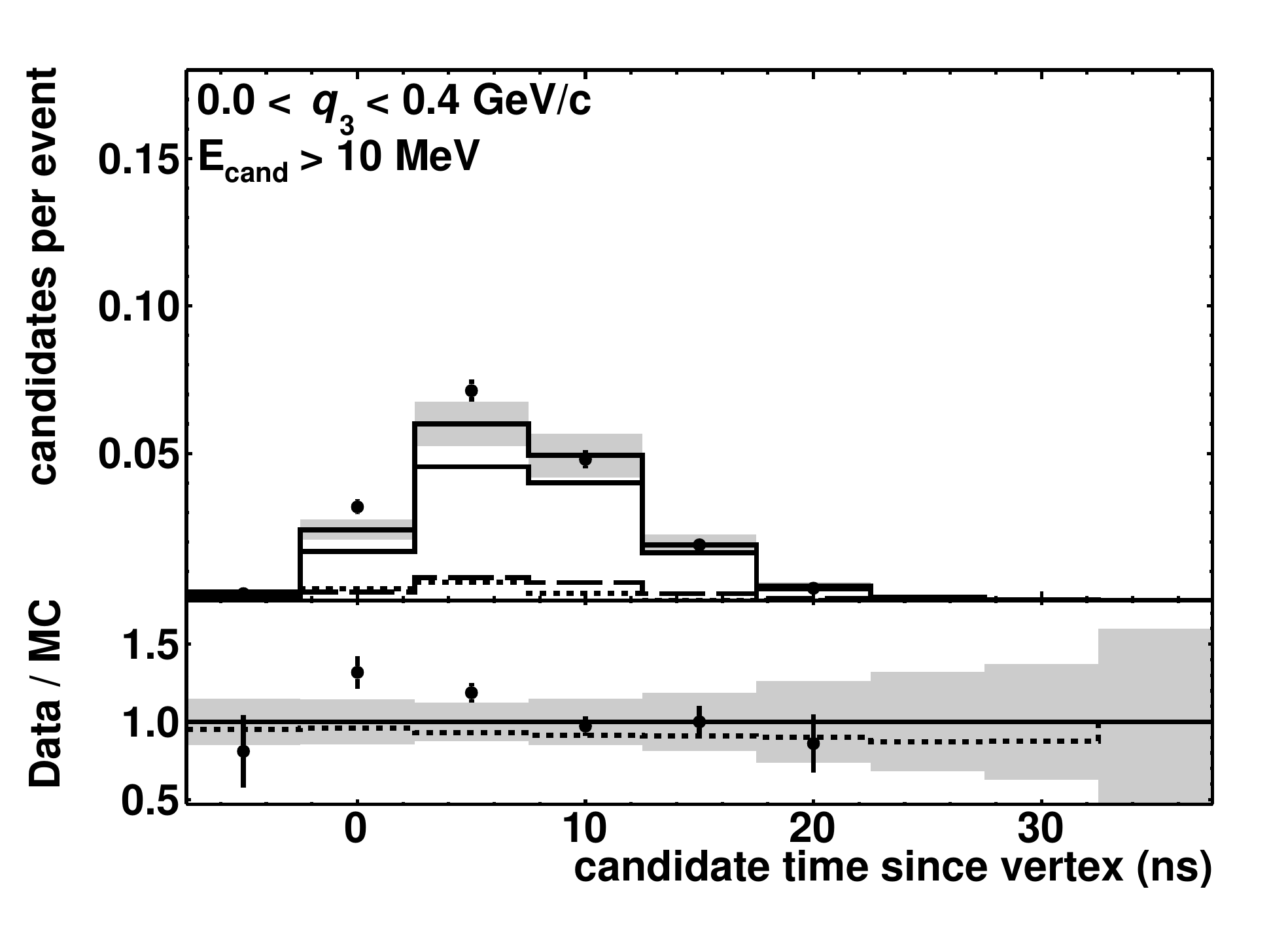}
\includegraphics[trim=0 0 0 30,clip,width=7cm]{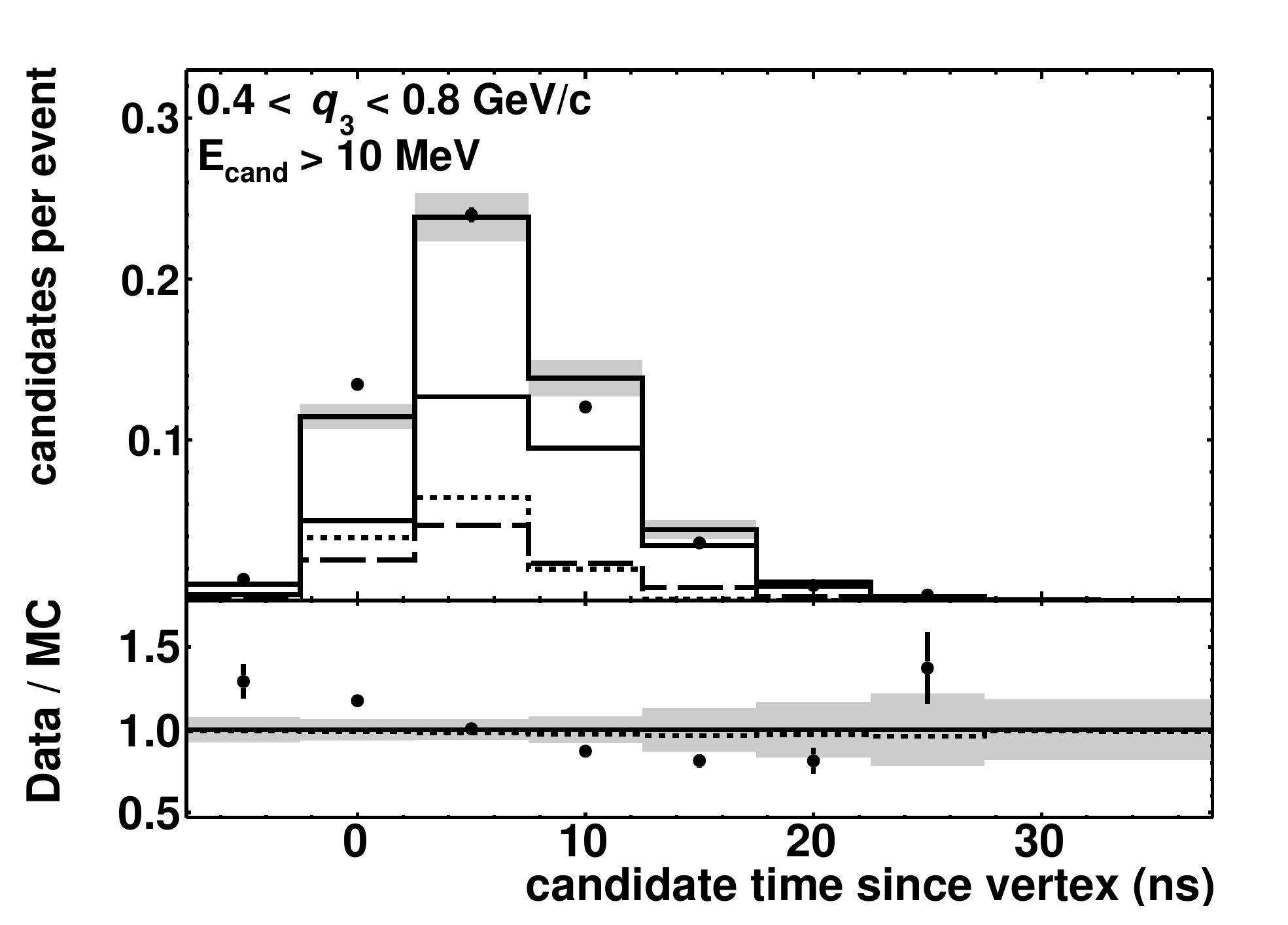}
\caption{Time of the candidate relative to the time of the
  interaction.  Data are shown with statistical uncertainties only; the simulation is shown with systematic uncertainties.
  Neutron candidates with energy
  deposits less than 10 MeV are
  shown for both ranges of $q_3$ in the upper plots, and higher-energy candidates are
  the lower plots.   Bins with very large data statistical
  uncertainties are not shown.
\label{fig:time}}
\end{center}
\end{figure*}

\subsection{Time-of-flight distribution}

The time difference between the neutron candidate and the interaction
time, $t_n - t_0$ shown in Fig.~\ref{fig:time} produces separation of the prompt electromagnetic component
from the slowest neutron component. 
Because the simulated overprediction is so prominent in the first two bins, 
the following distributions will be separated into subsamples below 
and above 10 MeV of candidate energy deposit $E_\mathrm{dep}$.    
The simulation's overprediction for candidates with reconstructed
energy less than 10 MeV in
the upper panels 
appears roughly uniform across all high-statistics bins of time-of-flight.
For higher-energy candidates in the lower panels there are trends
beyond the edge of the error band to relatively overpredict the latest times, either
from neutrons that traveled either the farthest or the slowest.
The modified G{\small EANT}4 response model describes the data better than the
modified {\small GENIE} model for the $0.4 < q_3 < 0.8$ GeV/c top-right panel.
By construction the G{\small EANT}4 benchmark has no effect on the lower panels,
though the G{\small EANT}4 cross section uncertainties are significant in the
error band.  The {\small GENIE}
modification reduces the slowest neutrons in the sample and has a slight shape effect smaller than the predicted
error band.

Fluctuations to negative times are compatible with the
time resolution of 4.5 ns for single-cluster candidates
(shown later in Fig.~\ref{fig:resolutions}): about one bin in these histograms.
Systematic uncertainties directly from the measurement of time of
flight for an individual event contribute negligibly.  The 
simulation of the timing distribution is taken from a separate, {\it in situ}
muon sample.  The lack of bias is independently confirmed
using clusters on the muon tracks of interactions in the selected
sample and reconstructed the same as neutron clusters.

\begin{figure*}[t]
\begin{center}
\includegraphics[trim=0 15 0 0,clip,width=7cm]{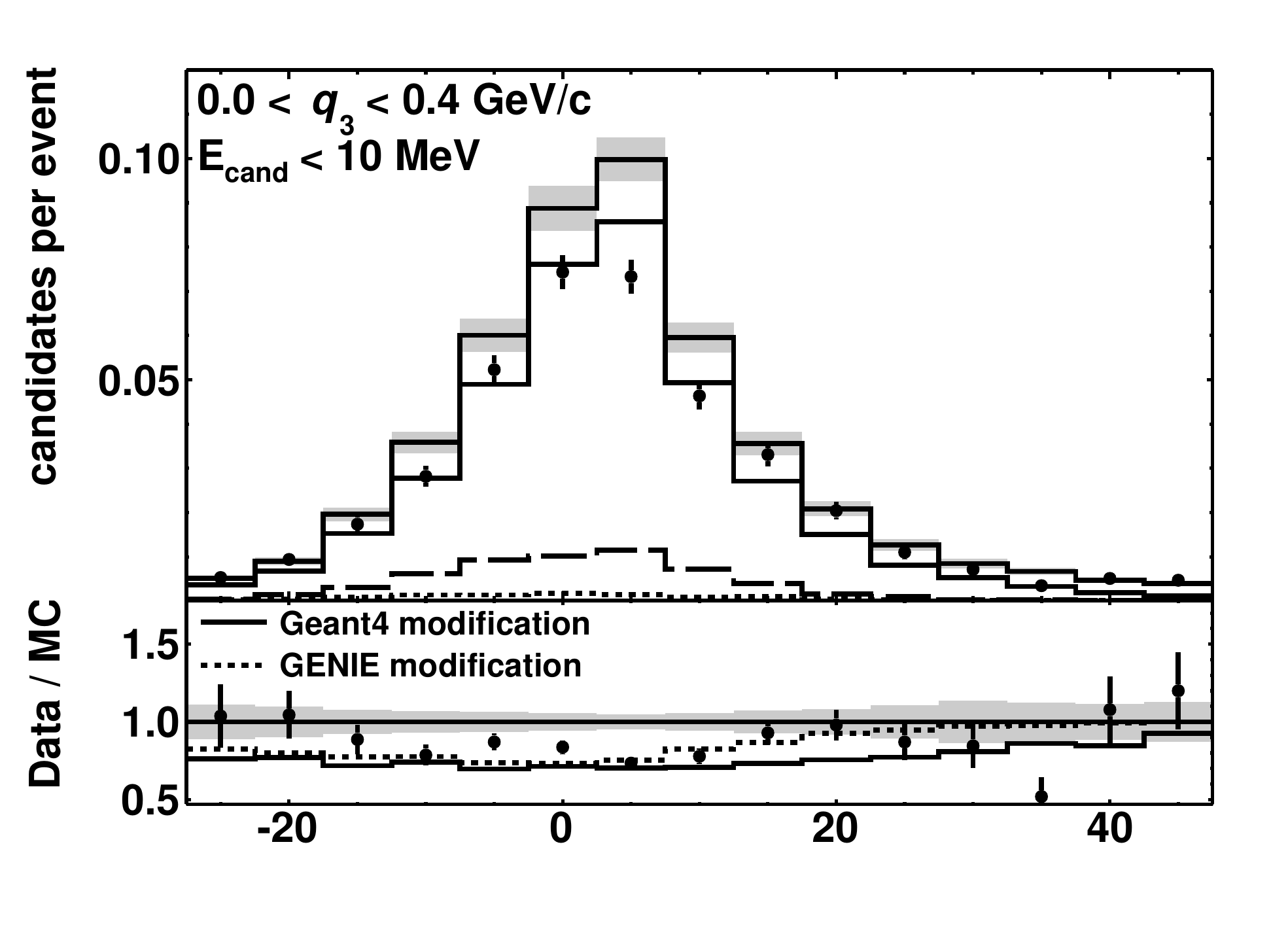}
\includegraphics[trim=0 15 0 0,clip,width=7cm]{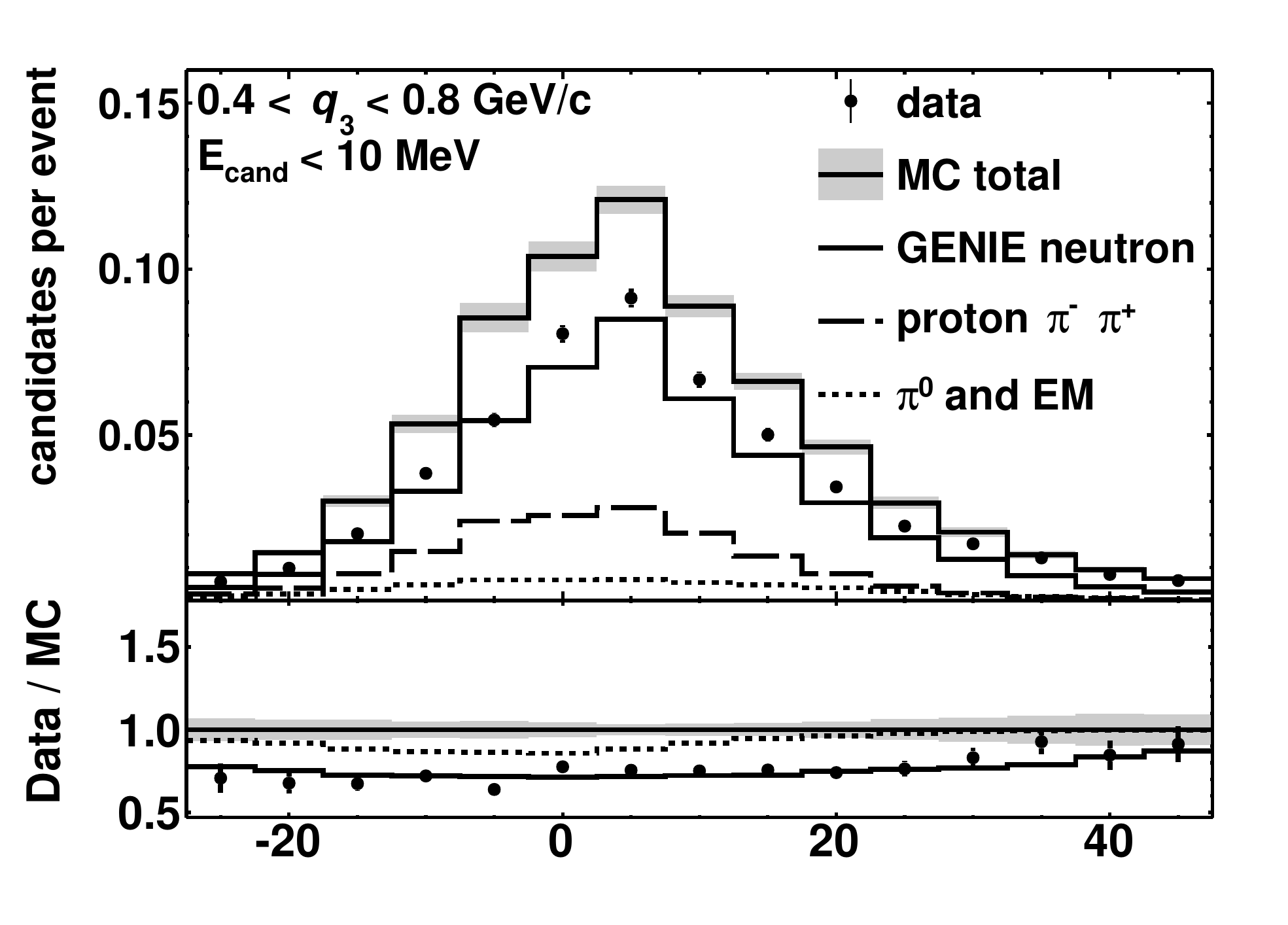}
\includegraphics[trim=0 0 0 30,clip,width=7cm]{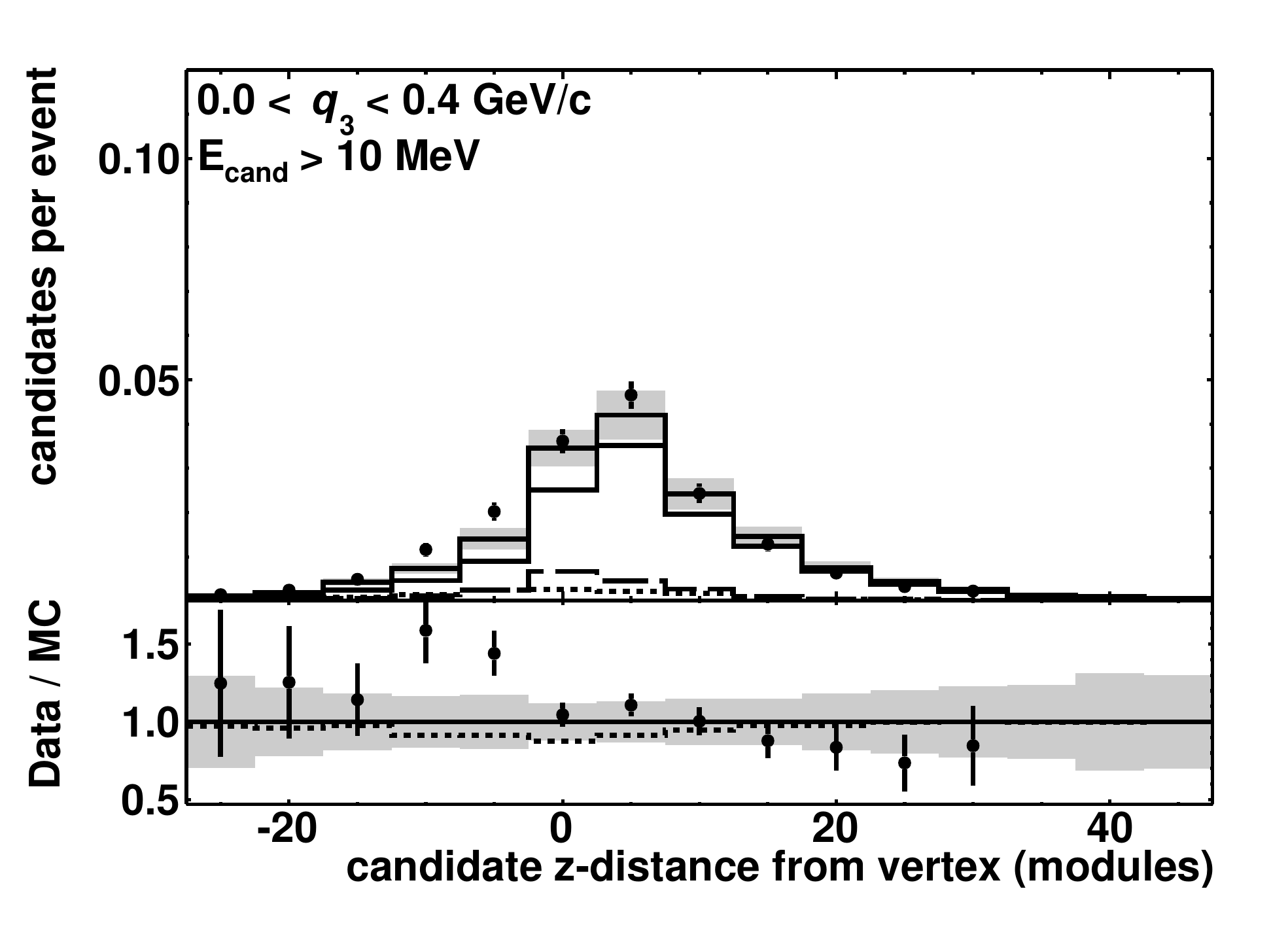}
\includegraphics[trim=0 0 0 30,clip,width=7cm]{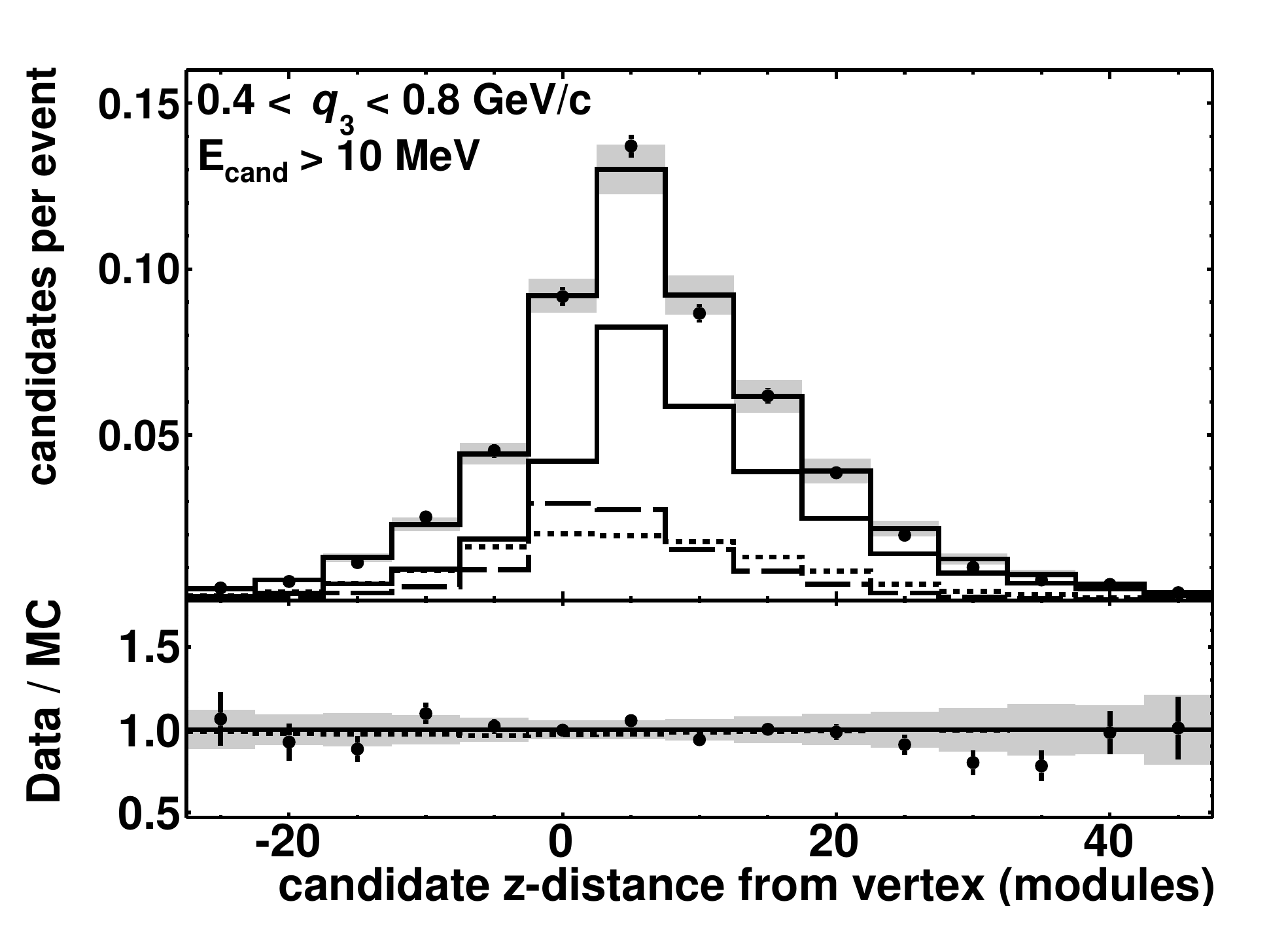}
\caption{Position of the candidate relative to the interaction point in the upstream/downstream
  direction.   Data are shown with statistical uncertainties only; the simulation is shown with systematic uncertainties.
  Neutron candidates with energy
  deposits $<$10 MeV are shown for 
  both ranges of $q_3$ in the upper plots, and higher-energy candidates are
  the lower plots.   Bins with very large data statistical
  uncertainties are not shown.
\label{fig:zposition}}
\end{center}
\end{figure*}

The same systematics described in the previous section are evaluated
for this distribution.
The G{\small EANT}4 neutron cross section
model uncertainties enhance or reduce the appearance of neutrons that travel
the farthest, and so have the
longest times.  It dominates the error bar in all bins beyond 15 ns.
The other uncertainties described previously contribute roughly equal
amounts in the center of the distribution.

\subsection{Position upstream or downstream}

The overprediction of candidates with energy deposits less than 10 MeV
appear broadly around the interaction point, shown in Fig.~\ref{fig:zposition}.  
In the $E_\mathrm{dep}<10$ MeV samples, the oversimulation may be 
prominent near the interaction point in the top left plot, while it
spans the detector for the higher momentum transfer sample in the top right plot.
For higher-energy candidates, the simulation does well overall except for two
underpredicted bins in the backward direction of the $q_3 < 0.4$ GeV/c
panel (lower left).

There are more neutron candidates in the forward direction, where the QE
process is especially relevant.
In contrast, candidates from any process
involving multiple particles can end up going backward from the
interaction point.   This includes
multibody reactions {\it 2p2h}, $\Delta$, and FSI in the nucleus, and neutrons produced when protons
and pions and neutrons reinteract in the detector.   

The cross
section for neutron reinteractions is the direct and dominant uncertainty in the
downstream region, as it was for long times.  
Again, the other uncertainties contribute roughly equally in the peak
of the distribution.

\begin{figure*}[t]
\begin{center}
\includegraphics[trim=0 15 0 0,clip,width=7cm]{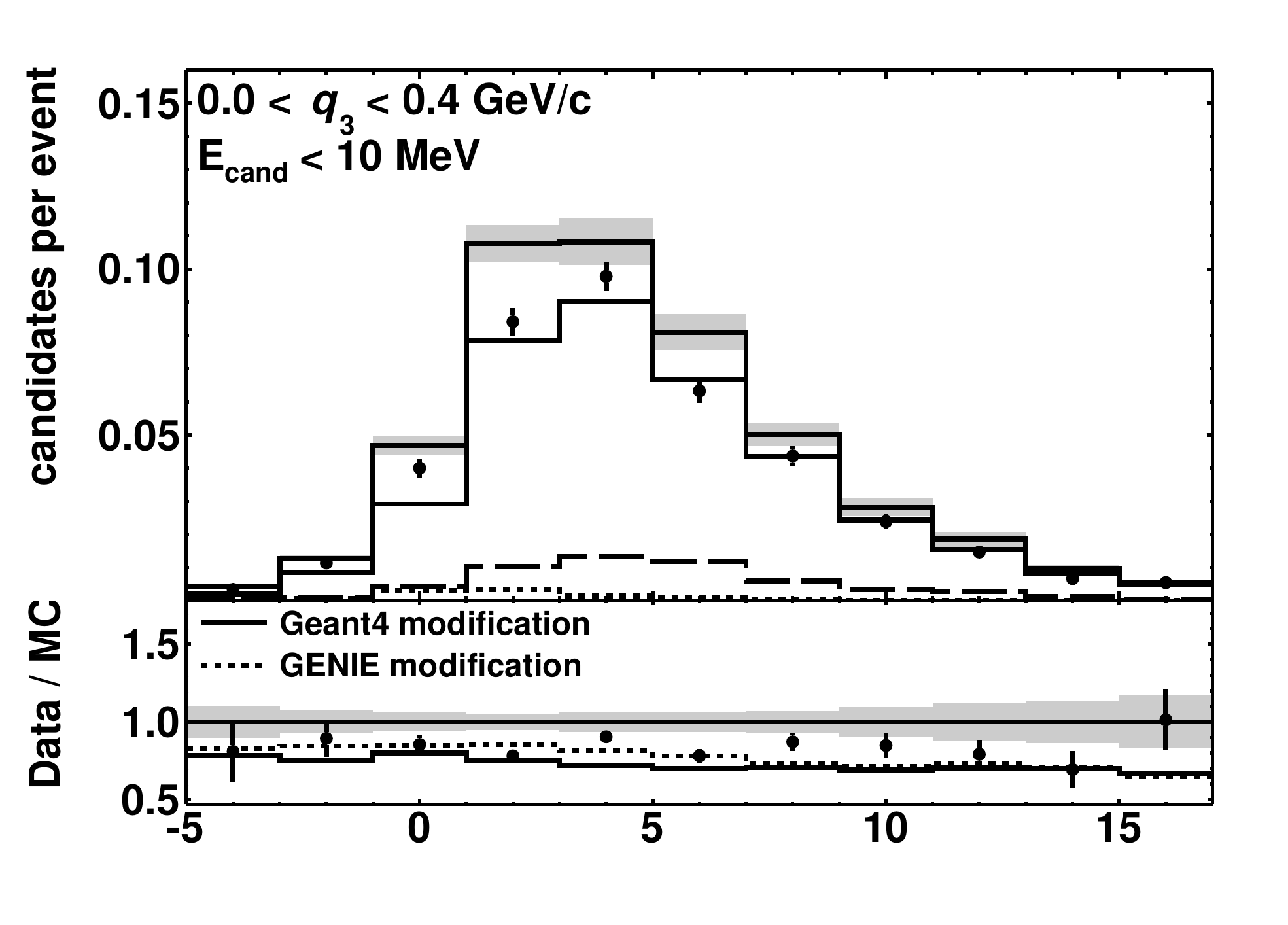}
\includegraphics[trim=0 15 0 0,clip,width=7cm]{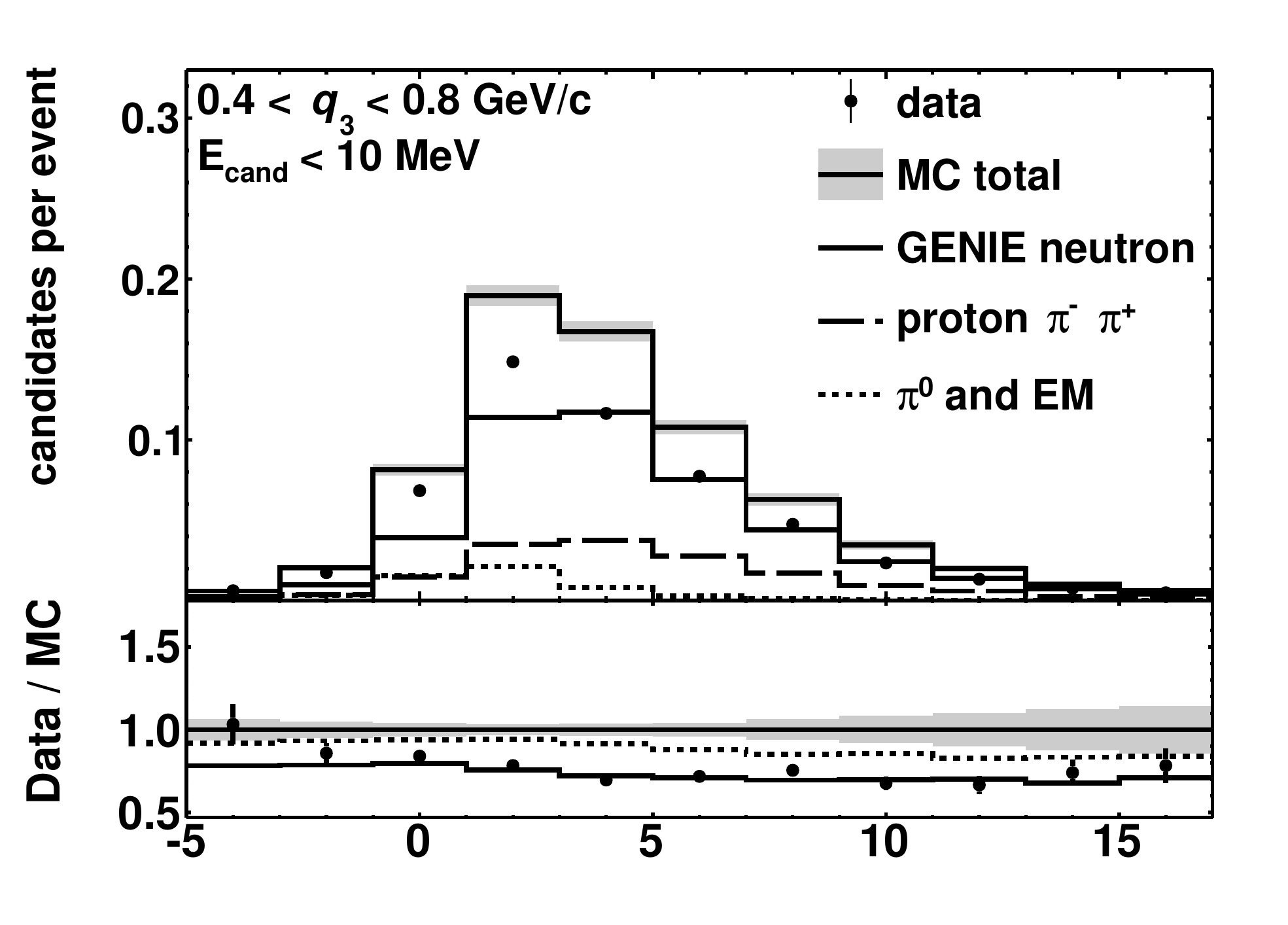}
\includegraphics[trim=0 0 0 30,clip,width=7cm]{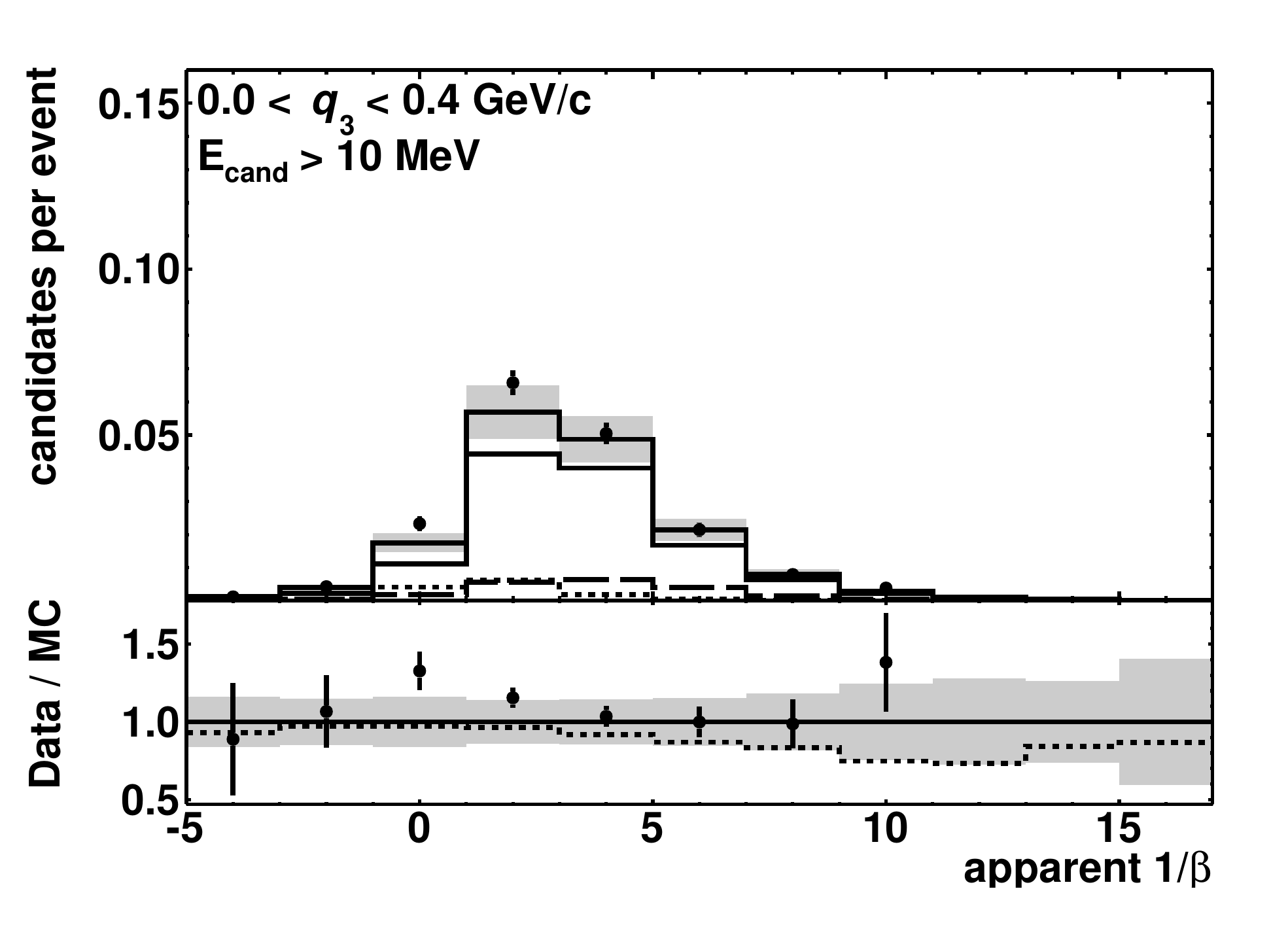}
\includegraphics[trim=0 0 0 30,clip,width=7cm]{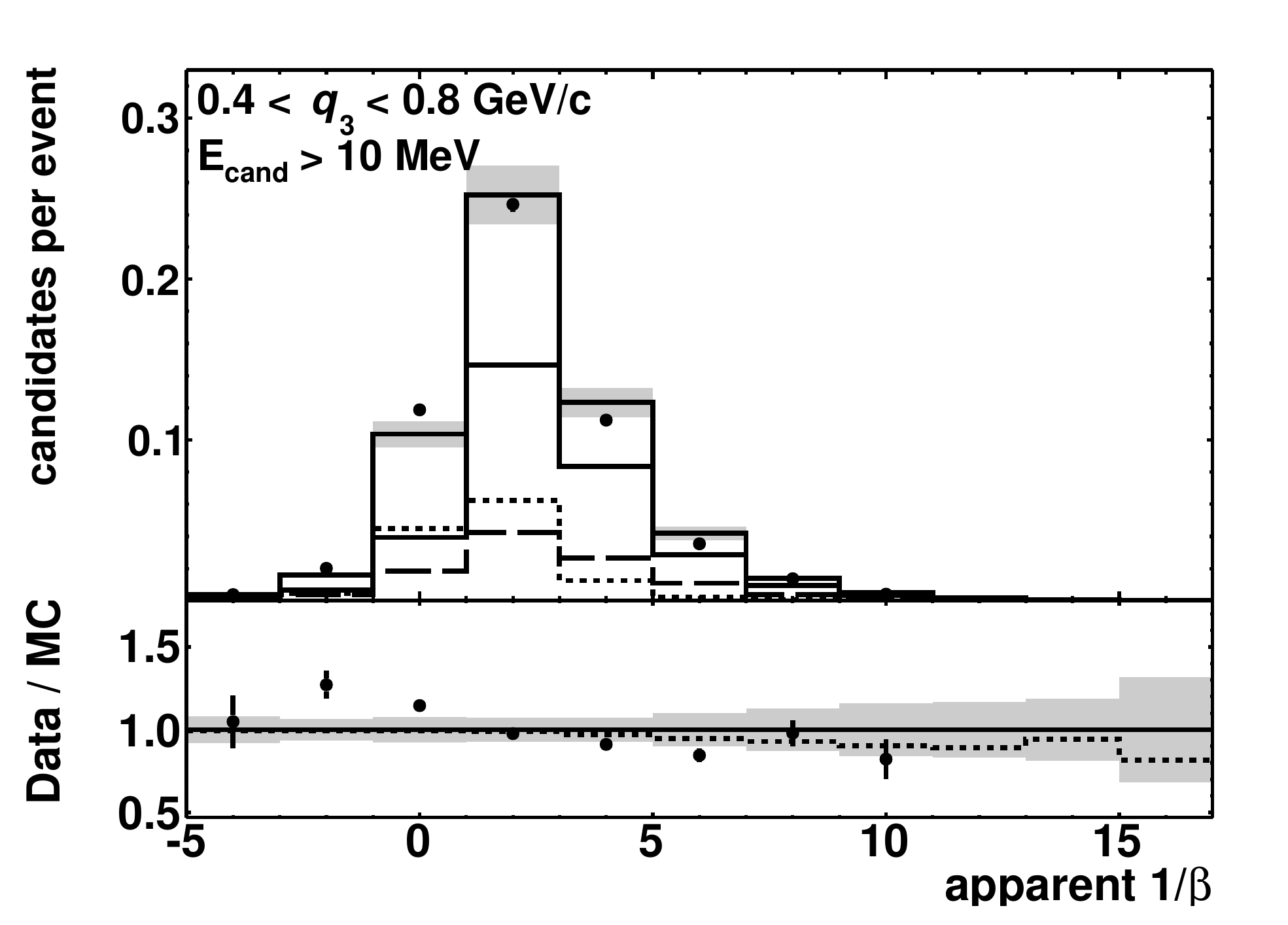}
\caption{Apparent $1/\beta$ of the particle causing the neutron 
  candidates, expressed as a fraction of the speed of 
  light.  Data are shown with statistical uncertainties only; the simulation is shown with systematic uncertainties.
  Bins with very large data statistical
  uncertainties are not shown.
\label{fig:oospeed}}
\end{center}
\end{figure*}

\subsection{Particle speed}

We estimate the apparent speed of the particle as a fraction of the speed of
light, $\beta$.   Because the timing resolution plays a crucial role,
and for better treatment of zero and negative times,
it is more clear to present 
$$\mathrm{apparent }\; 1/\beta = \mathrm{speed~of~light} \times \mathrm{time} / \mathrm{2D~distance.}$$
The result is shown in Fig.~\ref{fig:oospeed}.

The estimate of a two-dimensional (2D) distance can be made of the hypotenuse of
the distance in Z shown Fig.~\ref{fig:zposition} and the distance
along the one X, U, or V transverse direction measured for
single-cluster candidates.
This distance is transverse from the neutrino's path, not from the muon.
If the neutron candidate is made
of activity in more than one plane, the longest transverse position is
used.  This distance estimator is necessarily smaller than the true
distance the neutron traveled, because it is missing the third of
three coordinates, and because some neutrons bounce and take an
indirect path to the point where an energy deposit is observed.
This distribution has properties similar to the one in
Fig.~\ref{fig:zposition}, and is not shown.  

\begin{figure}[htb]
\begin{center}
\includegraphics[width=7cm]{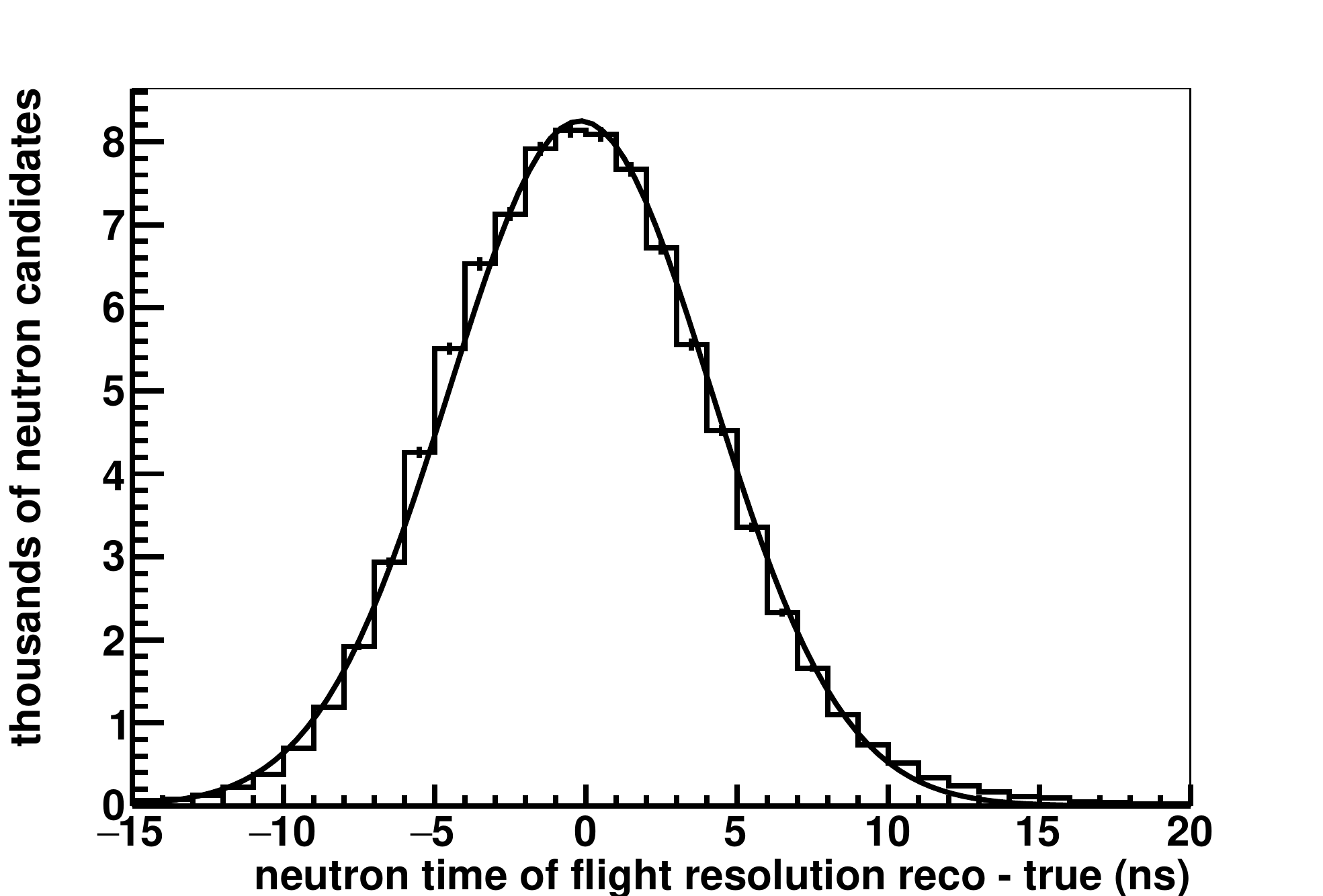}
\includegraphics[width=7cm]{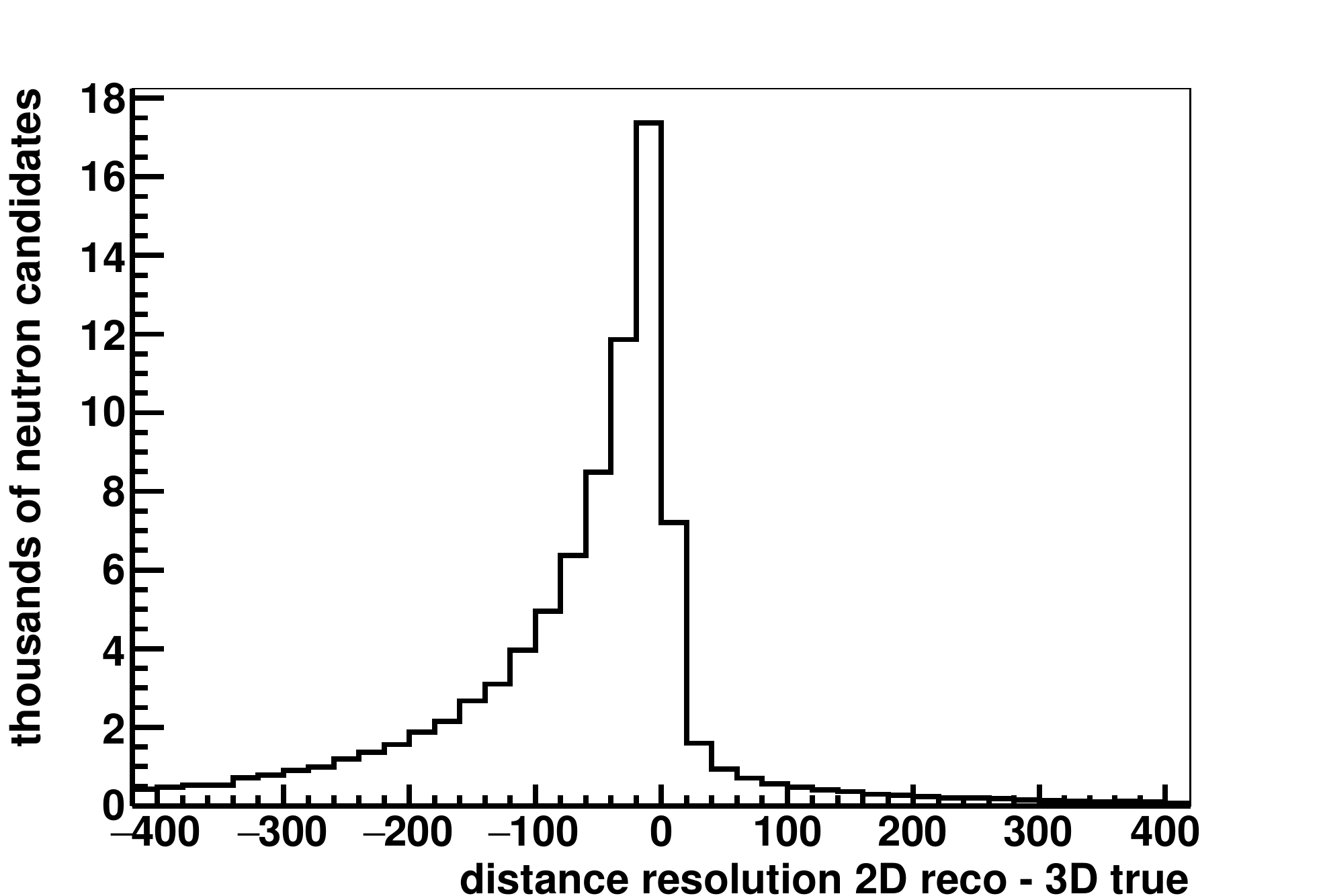}
\caption{Detector resolutions on the time and distance inputs to
  $1/\beta$ from the simulation, only for neutron candidates of which the
  origin was a neutron from the {\small GENIE} simulation.  There is no timing bias and the Gaussian fit to the timing
  resolution has $\sigma = 4.5$ ns.  At the speed of light, the mean
  and rms of the distance
  distribution correspond to 0.15 and 0.3 ns.  Neutron multiple
  scattering effects are not included in the lower plot.
\label{fig:resolutions}}
\end{center}
\end{figure}

The systematic underestimate of the 2D distance
means a systematic overestimate of
$1/\beta$ of about 0.8 and a rms resolution between 2 and 3, driven
largely by the timing resolution.   The resolution for the slowest
particles with true $1/\beta > 5$ is the worst because they
do not travel very far and are observed closest to the interaction
point.  They have a resolution of around 4 and a bias of -0.8.
The detector-only (without the effect of neutron multiple scatter)
time and 2D distance resolutions are shown in
Fig.~\ref{fig:resolutions}.
For neutrons, $1/\beta=5$ implies 20 MeV and $1/\beta=10$ implies 5 MeV; however
the latter are expected to rarely produce candidates (see again Fig.~\ref{fig:efficiency})
and the population beyond $1/\beta = 10$ must be from fluctuations in
the assigned time and distance.
The resolutions and thresholds are such that the apparent $1/\beta$ 
is not usefully transformed into a kinetic energy distribution.

The G{\small EANT}4 cross section uncertainty, prominent in the time
and z distance distributions separately, is much reduced and has
little shape.   A smaller or larger mean free path in the simulation affects both
the time and the distance simultaneously.   Other uncertainties
contribute similarly across these distributions.  The hadronic energy
scale and FSI uncertainties (migrations in $q_3$) are especially significant in
the first four bins of both the $E_{cand} > 10$ MeV (lower plots),
which is where discrepancies remain.

The electromagnetic component peaks near $1/\beta = 1.8$, shown as the
dotted line in the upper panels of each figure.
The neutron component peaks instead near $1/\beta \sim 4$ for
candidates from the lowest
kinetic energy neutrons and $1/\beta \sim 2.5$ for faster neutrons.
The {\small GENIE} benchmark modification produces a reduction in
the slowest neutron component that goes in the
direction of the data, but again does not match the overall magnitude
of the discrepancy in the upper right panel.

\subsection{Neutron multiplicity}

\begin{figure*}[tbh!]
\begin{center}
\includegraphics[trim=0 56 0 0,clip,width=8cm]{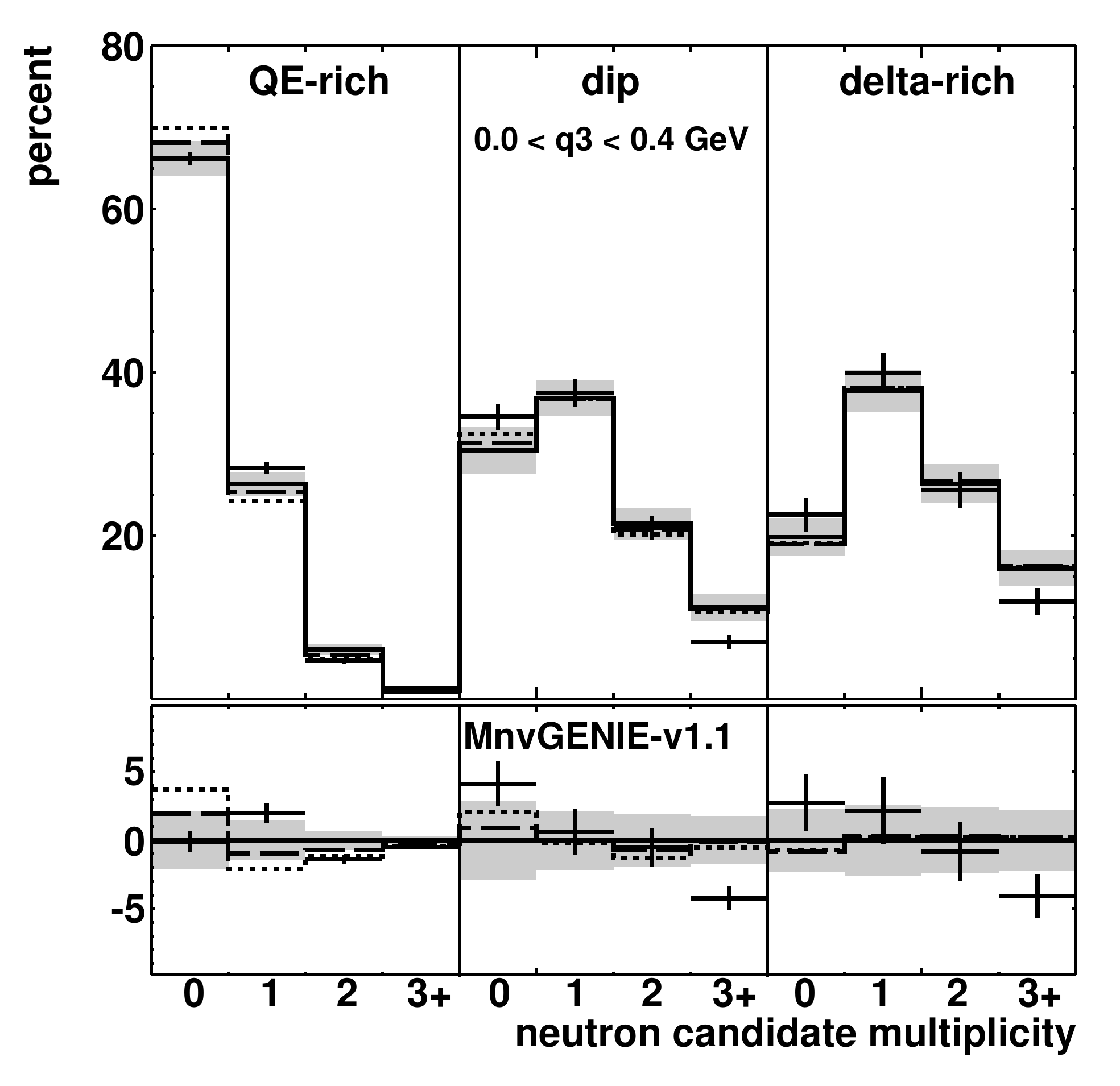}
\includegraphics[trim=0 56 0 0,clip,width=8cm]{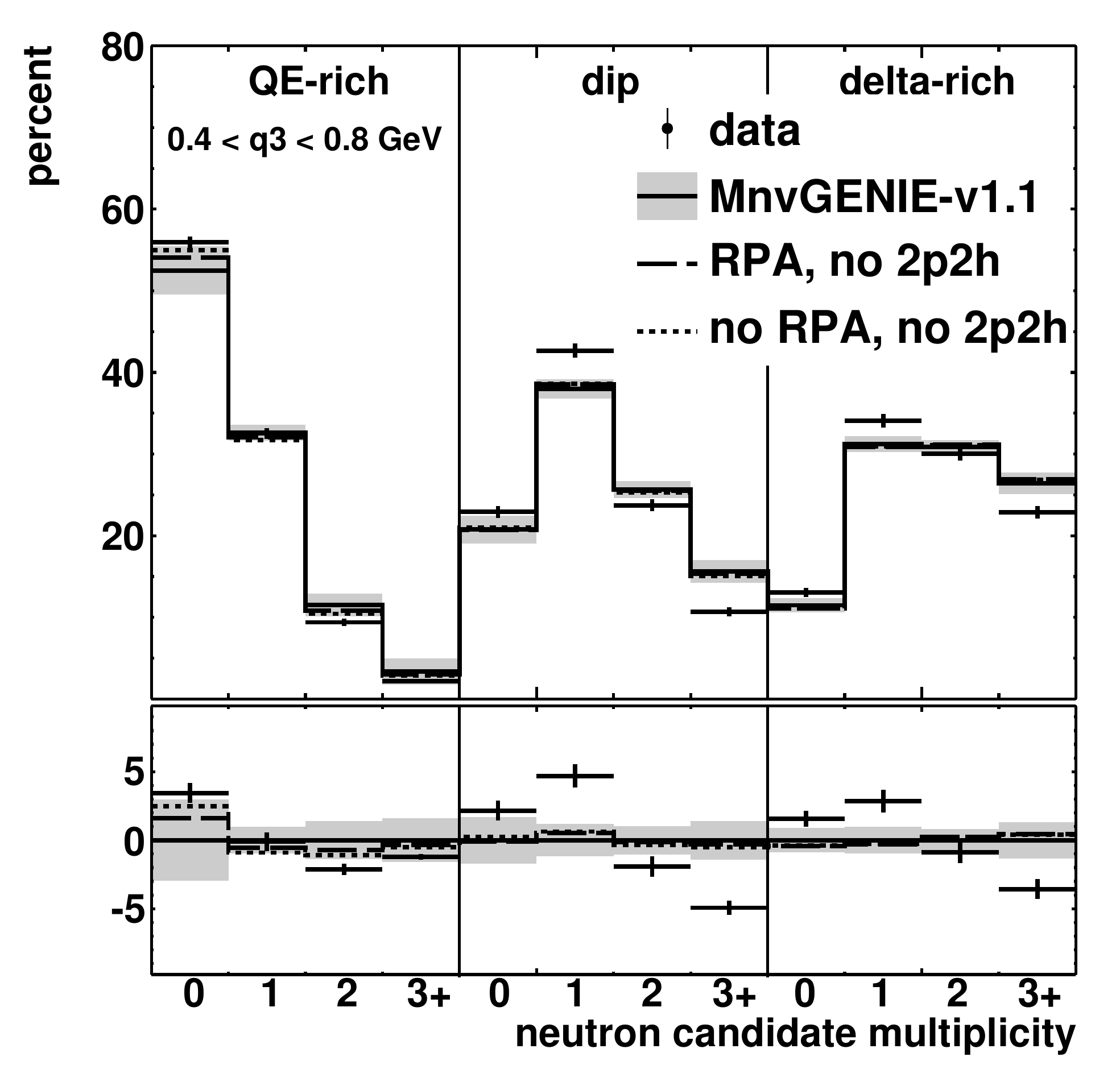}
\includegraphics[trim=0 56 0 360,clip,width=8cm]{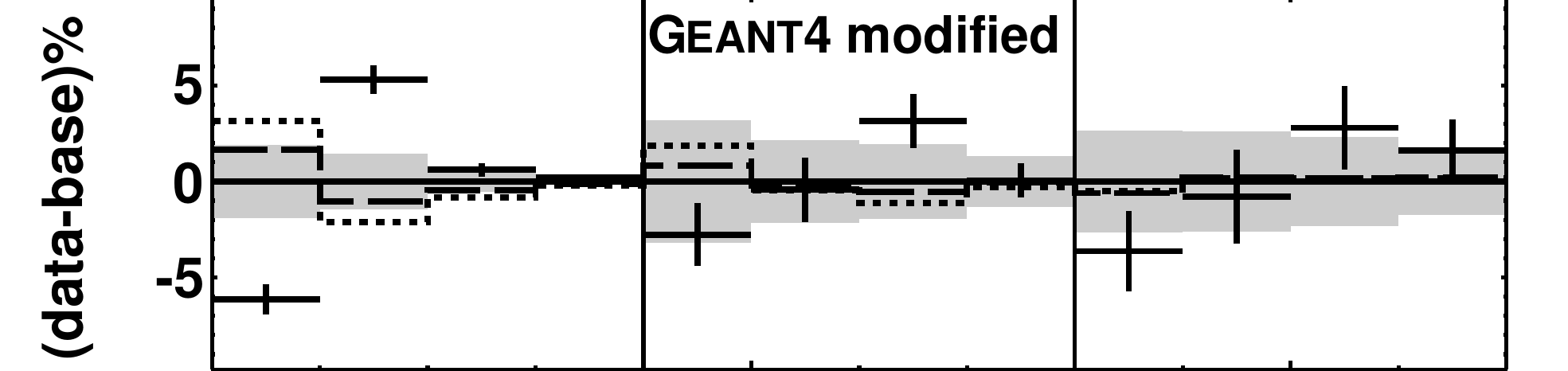}
\includegraphics[trim=0 56 0 360,clip,width=8cm]{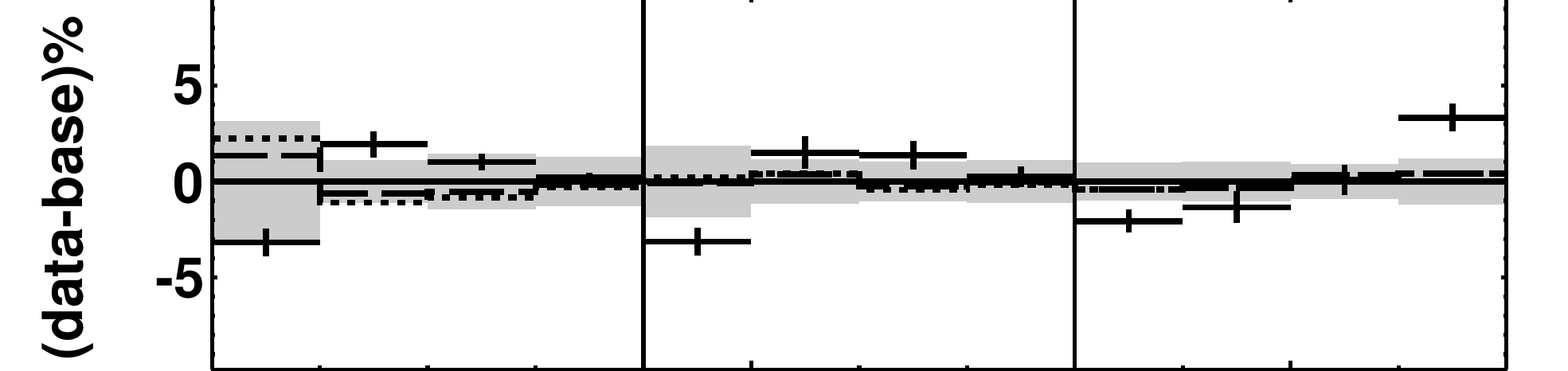}
\includegraphics[trim=0 15 0 360,clip,width=8cm]{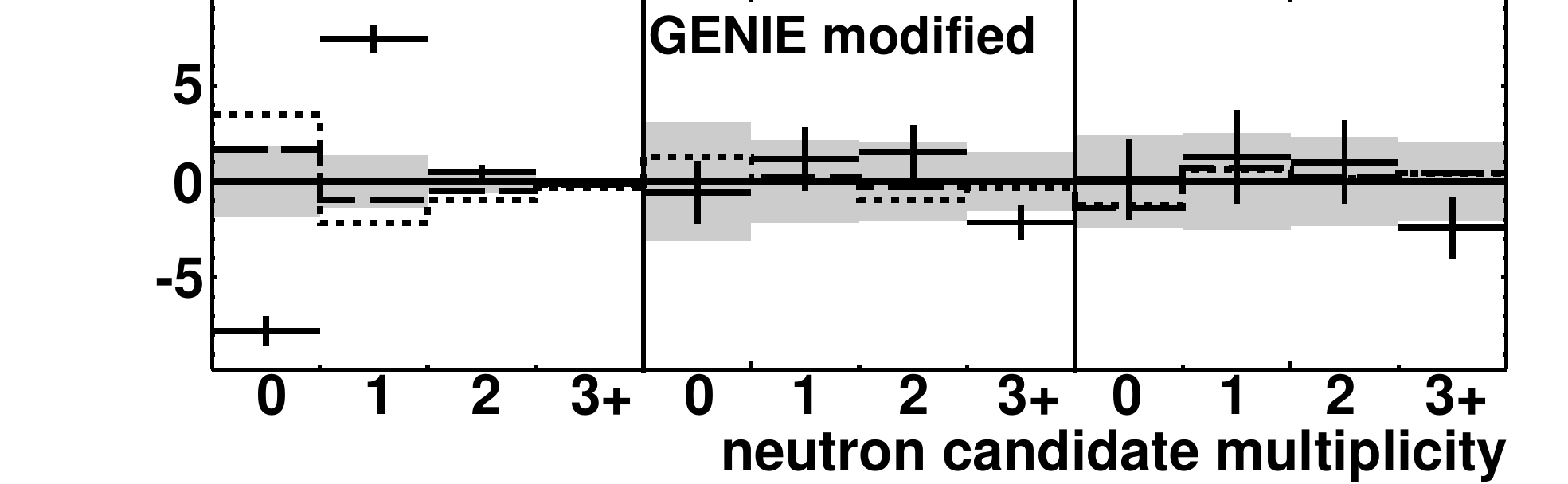}
\includegraphics[trim=0 15 0 360,clip,width=8cm]{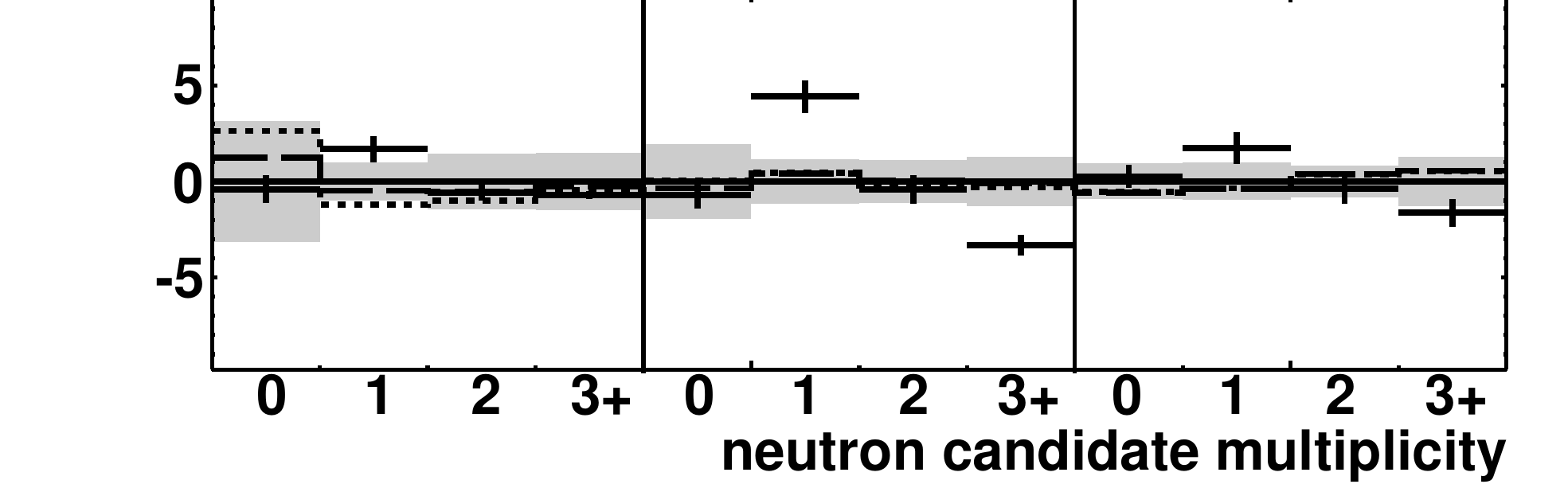}
\caption{Candidate multiplicity
  distribution for all six subsamples, $0 < q_3 < 0.4$ (left) and $0.4
  < q_3 < 0.8$ GeV/c (right), with subpanels for the QE-rich, dip, and
  $\Delta$-rich regions. The top plot shows the reference MnvGENIE-v1.1
  simulation with a solid line and error band, and two variations that turn off completely the
  {\it 2p2h} component and then also turn off the RPA component.  The next row
shows the difference from the reference simulation.   The middle (lower) row
  of difference plots uses the modified G{\small EANT}4 benchmark
  (modified {\small GENIE}
  benchmark) for all distributions.
\label{fig:multiplicity}}
\end{center}
\end{figure*}

The overprediction of neutron candidates per event in the simulation also distorts the
multiplicity of candidates per event, shown as a percent of the total in Fig.~\ref{fig:multiplicity}.  
The difference in percent (not the
ratio) between the data and the reference simulation is shown in the first panel below
the main distributions for compact comparison.
The overprediction of neutrons masks the presence of
the {\it 2p2h} component and RPA screening because both also enhance the
percent of events with neutron candidates.   Again, the {\small GENIE} and G{\small EANT}4
serve as benchmark
modifications showing the (modified simulation~-~reference) in percent in the lower
two rows, but now to expose these multinucleon features of the data.
The two regions of $q_3$ shown in the previous figures
are subdivided according to hadronic energy
to produce distributions for QE-rich, dip, and $\Delta$-rich
subsamples, six in total.  
In the top panels the oversimulation of neutron candidates is most evident
in the dip and $\Delta$-rich subsamples where the simulation significantly
overpredicts how many events have three or more neutron candidates and
underpredicts how many have none.  

The QE-rich subsets uniquely offer predicted sensitivity to
multinucleon effects.   The long dashed line in
Fig.~\ref{fig:multiplicity} completely removes the simulated {\it 2p2h}
component.  These events preferentially had multiple neutrons in the
first place, so removing them increases the bin with zero candidates.
The short-dashed line shows that removing also the QE RPA
screening effect adds back events of which the outgoing neutron was lower
energy and less likely to make a candidate, also increasing the bin
with zero candidates.  In contrast to the QE-rich panels, 
there is no sensitivity to these
multinucleon effects in the other panels;
all reactions produce similar numbers of neutrons after FSI.
Variations of RPA and {\it 2p2h} processes are the same size as the
uncertainty bands in those panels.

A different way to summarize the subdivision of the data:  
neutrino model details in Fig.~\ref{fig:multiplicity} are orthogonal to
the neutron details in the previous figures.   Modifications to the QE
and {\it 2p2h} models show up in QE-rich region here while the excess of
neutrons distorts all six panels similarly.   The opposite happens in
the previous figures; distortions of the spectra due to neutron
production details are evident, but modifications to the {\it 2p2h} and QE
models are largely flat with neutron candidate time, position, and
speed. 

What is desirable is to tune the neutron model to the dip and $\Delta$-rich regions, a
common technique when there are sidebands to a signal selection.
Such a tune would correct and constrain the mismodeled neutron effects in the multinucleon
sensitive distributions.  Though we do not directly have tunable
parameters, a simplified version is obtained by remaking the
distributions while applying the benchmark
modifications to G{\small EANT}4 (third row) and {\small GENIE} (bottom row).
The resulting dip and $\Delta$-rich regions are now consistent with
uncertainties for $0 < q_3 < 0.4$ GeV/c.   The G{\small EANT}4 modification
produces better distributions for $0.4 < q_3 < 0.8$ GeV/c, perhaps overcorrecting,  while
the {\small GENIE} modification produces mild improvement that does not go far
enough.  
Both roughly mimic the behavior of these benchmarks in the
energy, time, position, and speed distributions.

The benchmark modifications also reduce the preponderance of simulated
neutron candidates in the QE-rich signal region.  This enables further
interpretation of the presence of multinucleon effects and other
unsimulated processes.   
Especially in the leftmost two bins in the lower two rows of panels,
the modified simulations now have a 6\% to 8\% 
underprediction of events with one neutron candidate and an
overprediction of events with none.  This is roughly two times the
combined effects of our current RPA, {\it 2p2h} multinucleon models and all
systematic uncertainties.
In both cases,
the resulting new predictions hint the data want even more {\it 2p2h}
interactions or RPA screening than the reference MnvGENIE-v1.1
simulation.   

Sideband tuning usually takes a poor model and improves it before
extracting the physics quantities of primary interest.  
In this case, it takes what was naively a reasonable description in
the QE-rich sample and
indicates disagreement beyond the available multinucleon effect
models.   However, all the {\it relative} trends (not shown) of the whole sample in energy,
distance, time, and speed shown in previous figures are also present
for the QE-rich sample, suggesting that the sideband adjustment will
succeed and the new disagreement is a robust observation.
Alternatives to simply adding more {\it 2p2h} or RPA screening that
would correct the resulting model disagreement are to add an additional process like
deexcitation photons from carbon, or a
more nuanced, QE-specific version of either of the benchmark modifications.

In all the previous distributions, the current error bands preclude
further detailed tests of the magnitude of RPA, the relative proton+neutron and
neutron+neutron content of the {\it 2p2h} process, and the need for a low
$Q^2$ suppression of resonances \cite{Adamson:2014pgc}.   
The sensitivity would be limited even if there were no large
discrepancy, but modified RPA, {\it 2p2h}, or resonances
would not explain the $E_\mathrm{dep}<10$ in the whole sample nor in
the QE-rich sample on its own.

Prior MINERvA measurements show distributions with sensitivity to the RPA and {\it 2p2h}
multinucleon models that may also be sensitive to neutron effects.   
In Fig.~3 of Ref.~\cite{Gran:2018fxa} the reconstructed
hadronic energy for this same sample is improved with
the addition of RPA and a tuning of {\it 2p2h} to the neutrino data in
Ref.~\cite{Rodrigues:2015hik}, but the antineutrino agreement is not perfect.  
The untracked energy within 100~mm from the interaction point of antineutrino reactions
in Fig.~25 of Ref.~\cite{Patrick:2018gvi} is effectively in the excluded
region of this analysis.   That distribution is also not as well described by
MnvGENIE-v1 compared to the equivalent Figs.~35-36 of
Ref.~\cite{Ruterbories:2018gub}.  In both examples, and based on the neutron observations in
this paper, the description of the
antineutrino data could be improved with reduction of the neutron component of the
reconstructed energy in the simulation, while having a
smaller effect on the neutrino-mode data.   Such a mechanism supposes
the reduced neutron energy goes missing rather than being offset by
additional charged hadron energy.

\subsection{Protons}

This final study is the analog to the untracked protons reported for the related neutrino
interaction sample of Ref.~\cite{Rodrigues:2015hik}, where the proton
multiplicity with an untuned Valencia {\it 2p2h} process is at the edge of the error
band of a prediction without it.   Protons are counted by observing single strips with at least
20 MeV near the interaction point, indicating the Bragg peak at the
end of the proton range.  Because the single strip could be at the
interaction point itself, the threshold is effectively just above 20
MeV.  T2K has recently presented results for proton multiplicity
\cite{Abe:2018pwo} using a tracking threshold kinetic energy of 100 MeV

\begin{figure*}[tbh!]
\begin{center}
\includegraphics[width=8cm]{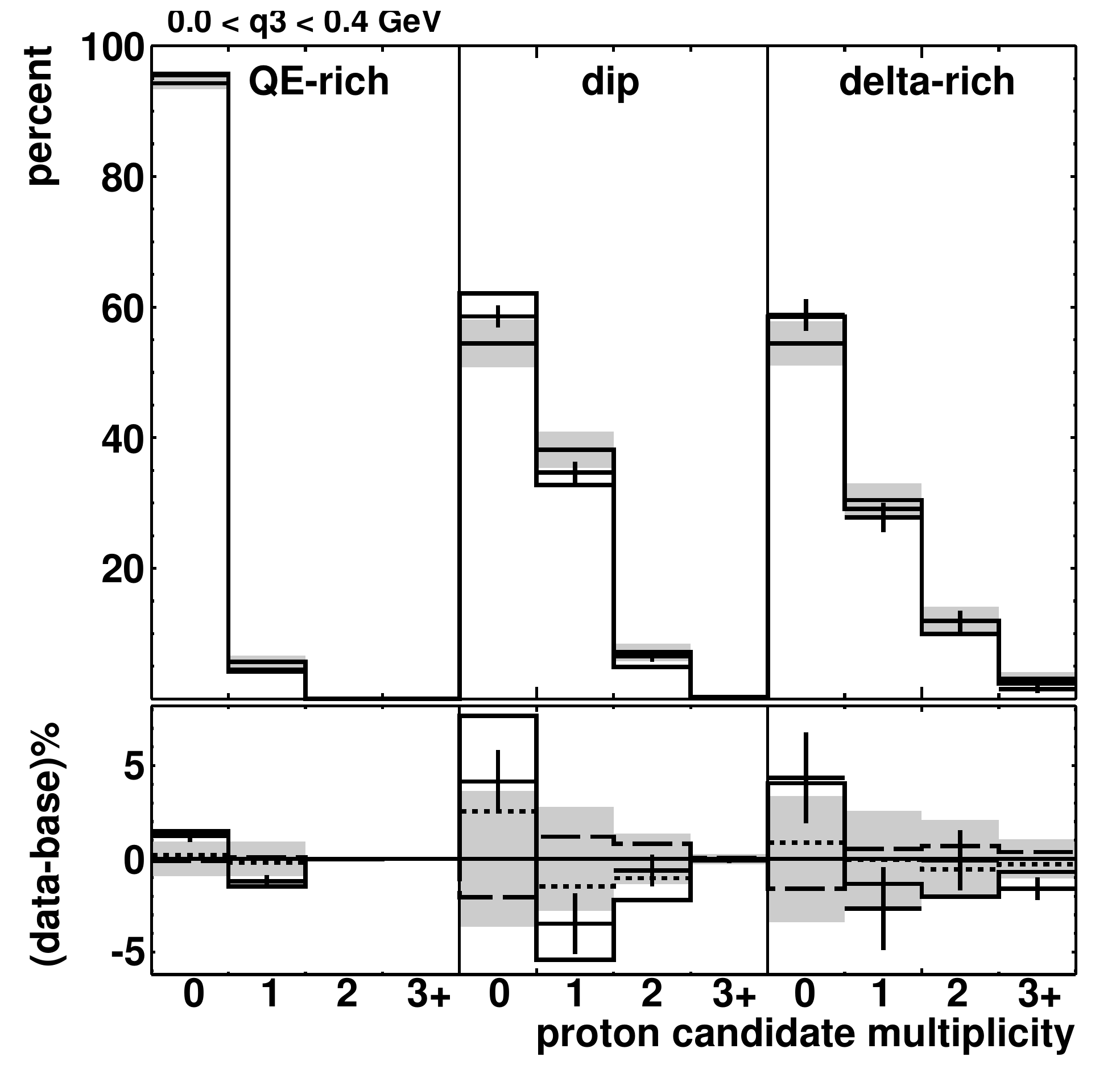}
\includegraphics[width=8cm]{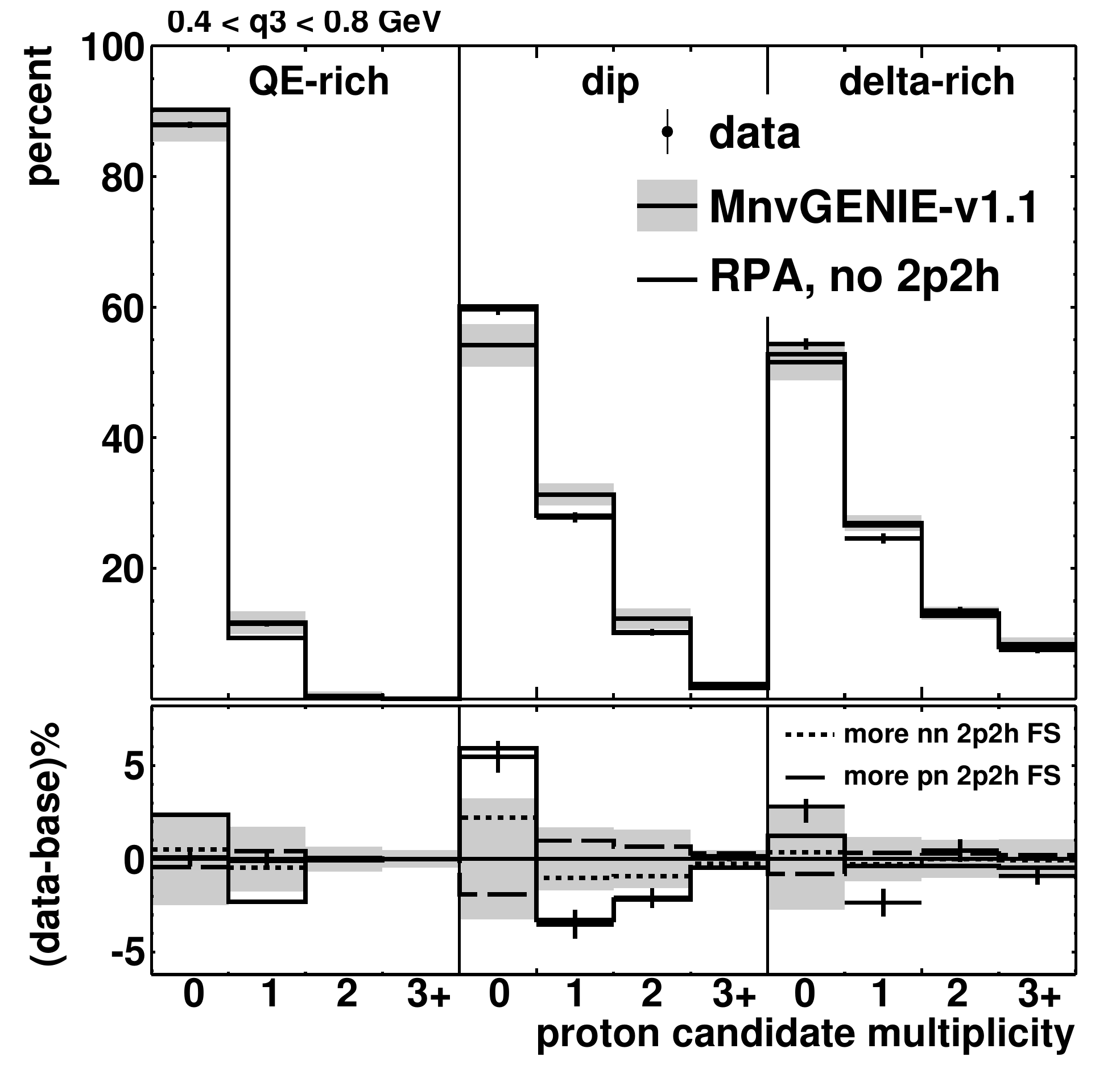}
\caption{Candidate proton multiplicity
  distribution for all six subsamples, $0 < q_3 < 0.4$ (left) and $0.4
  < q_3 < 0.8$ GeV/c (right), with subpanels for the QE-rich, dip, and
  $\Delta$-rich regions. The top plot shows the reference MnvGENIE-v1.1
  simulation with a solid line and error band and one variation that turns off completely the
  {\it 2p2h} component.  The subpanels show the difference from the
  reference MnvGENIE-v1.1 for two additional {\it 2p2h} variations:  the dotted line
  enhances only the {\it pn} initial states which give {\it nn} final states (FS) and the dashed
  line enhances only the {\it pp} initial states leading to {\it pn} final states.
\label{fig:protmultiplicity}}
\end{center}
\end{figure*}

Protons are also more common in the final state of a {\it 2p2h} antineutrino reaction,
relative to antineutrino QE reactions.   Repeating that earlier
strategy reveals marginal sensitivity, shown in
Fig.~\ref{fig:protmultiplicity}.   The reference simulation is shown
with a solid line and systematic error band, a simulation with no {\it 2p2h}
process at all is shown with the second solid line that has fewer
proton candidates and more zero-proton events.   The RPA screening
applied to the QE process has negligible effect on this proton distribution,
and an additional line without it is not included.
Two additional model variations are shown in the lower panels as the
difference from the reference simulation.   The dotted line enhances
(above the Valencia {\it 2p2h} model)
only the {\it pn} initial states which lead to {\it nn} final states
and predict a higher fraction of events with no protons.   The dashed
line enhances only the {\it pp} initial states which lead to {\it pn}
final states and produce more detectable protons.

Given the uncertainties, the proton multiplicity data are consistent with all model
variations presented.   Visually, the preference is for models that have
fewer protons, either from less {\it 2p2h} overall or from {\it 2p2h}
reactions that favor neutron only final states.   The latter is also
the conclusion from electron scattering results that indicate that
{\it pn} short-range correlated pairs are significantly more common
\cite{Subedi:2008zz} than like-nucleon pairs.
The {\small GENIE} FSI model uncertainties play the most significant
role in degrading the sensitivity, because they control how many additional protons are
ejected from the nucleus, especially for QE reactions.  This
sensitivity was not as strong in the neutrino case
\cite{Rodrigues:2015hik} where outgoing protons are the direct products
of the reaction.
Both FSI and
calibrated hadronic energy scale uncertainties have a significant
effect on the QE-rich panel where a higher-energy scale causes a
migration of events to the dip region or higher three-momentum
transfer panel.    Finally, the Birks' suppression uncertainty is also
significant throughout these distributions, its size is
half the total uncertainty shown.  It makes simulated protons
more or less likely to pass the 20 MeV selection.  Because {\it 2p2h}
variations are shown explicitly,
no uncertainty on the {\it 2p2h} process is included in the error band.


Liquid argon experiments
\cite{Acciarri:2014gev,Partyka:2015rua,Palamara:2016uqu,Adams:2018fud} have shown
more low-energy, charged
proton tracks in the simulation compared to data.  Under ideal
circumstances, this detector technology permits tracking of protons
with as little kinetic energy as 21 MeV. 
The {\small GENIE} model also produces more low-energy protons
than other neutrino event generators, correlated with its behavior for neutrons.
This supports that the {\small GENIE} benchmark modification may be part of
resolving these discrepancies.
In MINERvA, protons under 20 MeV would not meet the threshold for detection.
They would usually deposit all their energy in the same scintillator strip
in which the reaction occurred, and it takes 100 MeV before protons start to be
trackable.  So unlike neutrons and unlike protons in liquid argon detectors, multiple low-energy protons in a single
scintillator strip in MINERvA would be counted only once, if at all.

\section{Conclusion}

We have obtained the first time of flight, spatial, and speed (1/$\beta$) distributions of
neutrons from antineutrino interactions.   The reference simulation,
the components of which are widely used by neutrino oscillation experiments,
overestimates the number of neutron candidates by 15\% overall but by 25\%
for energy deposits less than 10 MeV, shown in the upper figures of Figs.~6-9.
A possible
interpretation is that the {\small GENIE} neutrino event generator and its FSI model
are overproducing the lowest-energy neutrons.  Also likely, the
G{\small EANT}4 and  detector models turn too many
neutron interactions into measurable activity.   Combinations and
variations of these two benchmark modifications are paths forward.  The discrepancy is
around two standard deviations from the combination of the other
sources of uncertainty.

Additional distortions may be present for candidates with energy
deposits more than 10 MeV.  The MC
overestimates long times of flight relative to short in Fig.~6, far
and forward relative to near and backward in Fig.~7,  and slow
relative to prompt in Fig.~8.   These discrepancies are just beyond
the error band, suggesting one or more of the uncertainties come close
to accounting for these data.

It is a reasonable assumption that similar overproduction of small energy
deposit neutron candidates is present for all subcomponents of the
sample.  In this case, the multiplicity distribution
in the QE-rich subsamples indicates
a preference for a model that has a combination of RPA
screening and a {\it 2p2h} component, both of which reduce the relative
proportion of events with zero neutron candidates.

  \begin{acknowledgments}

We are grateful to Juan Nieves, Ignacio Ruiz Simo, and Manuel Vicente Vacas for making their  RPA and 2p2h code 
available for study and incorporation into this analysis.
Samples of other neutrino event generator neutron predictions were prepared by Jake Calcutt, Luke Pickering, and Kendall Mahn using the NUISANCE framework.
This document was prepared by members of the MINERvA Collaboration using the resources of the
Fermi National Accelerator Laboratory, a U.S. Department of Energy, Office of Science, HEP User Facility.
Fermilab is managed by Fermi Research Alliance, LLC (FRA), acting under Contract
No. DE-AC02-07CH11359.
These resources included support for the \minerva construction project,
and support for construction also was
granted by the United States National Science Foundation under
Grant PHY-0619727 and by the University of Rochester. 
Support for scientists for this 
specific publication was granted by the United States National Science
Foundation under Grants PHY-1306944 and PHY-1607381.  
We are grateful for the United States National Science Foundation's
decade of direct support to the Soudan Underground Lab
outreach program, including Grant PHY-1212342; this analysis originated as the
research component for two summer undergraduate outreach interns.
Support for
participating scientists was provided by NSF and DOE (USA) by CAPES
and CNPq (Brazil), by CoNaCyT (Mexico), 
by Proyecto Basal FB 0821, CONICYT PIA ACT1413, Fondecyt 3170845 and 11130133 (Chile),
by CONCYTEC (Consejo Nacional de Ciencia, Tecnolog\'{i}a e Innovaci\'{o}n Tecnol\'{o}gica), 
DGI-PUCP (Direcci\'{o}n de Gesti\'{o}n de la Investigaci\'{o}n  - Pontificia Universidad Cat\'{o}lica del Peru), 
and VRI-UNI (Vice-Rectorate for Research of National University of Engineering) (Peru);
by the Latin American Center for Physics (CLAF);
and NCN Opus Grant No. 2016/21/B/ST2/01092 (Poland).
We thank the MINOS Collaboration for use of its
near detector data. Finally, we thank the staff of
Fermilab for support of the beamline, the detector, and the computing infrastructure.

\end{acknowledgments}

  \bibliographystyle{apsrev4-1}
  \bibliography{anuNeutronTag}



\end{document}